\documentclass[twocolumn]{aastex62}

\hypersetup{citecolor=blue, filecolor=blue, urlcolor=blue}

\usepackage{braket}
\usepackage{amsmath}
\usepackage{bm}
\usepackage{rotating}
\usepackage{comment}

\usepackage{lineno}

\newcommand{\lya}{Ly$\alpha\,$}

\newcommand{\cii}{[\ion{C}{2}]}

\newcommand{\mgii}{\ion{Mg}{2}}

\newcommand{\oiii}{[\ion{O}{3}]}

\revised{\today}
\submitjournal{ApJ}

\shorttitle{The Clustering of Luminous Quasars at $z\gtrsim6$}
\shortauthors{A.-C. Eilers et al.}

\begin{document}\sloppy\sloppypar\raggedbottom\frenchspacing
  
\title{\textbf{EIGER VI. The Correlation Function, Host Halo Mass and Duty Cycle \\ of Luminous Quasars at $z\gtrsim6$}}

\noindent
\author[0000-0003-2895-6218]{Anna-Christina Eilers}
\affiliation{Department of Physics, Massachusetts Institute of Technology, Cambridge, MA 02139, USA}
\affiliation{MIT Kavli Institute for Astrophysics and Space Research, Massachusetts Institute of Technology, Cambridge, MA 02139, USA}

\author[0000-0003-0417-385X]{Ruari Mackenzie}
\affiliation{Department of Physics, ETH Z{\"u}rich, Wolfgang-Pauli-Strasse 27, Z{\"u}rich, 8093, Switzerland}

\author[0000-0002-9712-0038]{Elia Pizzati}
\affiliation{Leiden Observatory, Leiden University, P.O. Box 9513, 2300 RA Leiden, The Netherlands}

\author[0000-0003-2871-127X]{Jorryt Matthee}
\affiliation{Department of Physics, ETH Z{\"u}rich, Wolfgang-Pauli-Strasse 27, Z{\"u}rich, 8093, Switzerland}
\affiliation{Institute of Science and Technology Austria (ISTA), Am Campus 1, 3400 Klosterneuburg, Austria}

\author[0000-0002-7054-4332]{Joseph F.\ Hennawi}
\affiliation{Leiden Observatory, Leiden University, P.O. Box 9513, 2300 RA Leiden, The Netherlands}
\affiliation{Department of Physics, University of California, Santa Barbara, CA 93106-9530, USA}

\author[0000-0002-4321-3538]{Haowen Zhang}
\affiliation{Department of Astronomy, University of Arizona, Tucson, AZ 85721, USA} 

\author[0000-0002-3120-7173]{Rongmon Bordoloi}
\affiliation{Department of Physics, North Carolina State University, Raleigh, 27695, North Carolina, USA}

\author[0000-0001-9044-1747]{Daichi Kashino}
\affiliation{National Astronomical Observatory of Japan, 2-21-1 Osawa, Mitaka, Tokyo 181-8588, Japan}

\author[0000-0002-6423-3597]{Simon J.\ Lilly}
\affiliation{Department of Physics, ETH Z{\"u}rich, Wolfgang-Pauli-Strasse 27, Z{\"u}rich, 8093, Switzerland}

\author[0000-0003-2895-6218]{Rohan P.\ Naidu}\thanks{NHFP Hubble Fellow}
\affiliation{MIT Kavli Institute for Astrophysics and Space Research, Massachusetts Institute of Technology, Cambridge, MA 02139, USA}

\author[0000-0003-3769-9559]{Robert A.\ Simcoe}
\affiliation{Department of Physics, Massachusetts Institute of Technology, Cambridge, MA 02139, USA}
\affiliation{MIT Kavli Institute for Astrophysics and Space Research, Massachusetts Institute of Technology, Cambridge, MA 02139, USA}

\author[0000-0002-5367-8021]{Minghao Yue}
\affiliation{MIT Kavli Institute for Astrophysics and Space Research, Massachusetts Institute of Technology, Cambridge, MA 02139, USA}

\author{Carlos S.\ Frenk}
\affiliation{Institute for Computational Cosmology, Department of Physics, University of Durham, South Road, Durham, DH1 3LE, UK}

\author{John C.\ Helly}
\affiliation{Institute for Computational Cosmology, Department of Physics, University of Durham, South Road, Durham, DH1 3LE, UK}

\author[0000-0002-2395-4902]{Matthieu Schaller}
\affiliation{Leiden Observatory, Leiden University, P.O. Box 9513, 2300 RA Leiden, The Netherlands}
\affiliation{Lorentz Institute for Theoretical Physics, Leiden University, PO Box 9506, NL-2300 RA Leiden, The Netherlands}

\author[0000-0002-0668-5560]{Joop Schaye}
\affiliation{Leiden Observatory, Leiden University, P.O. Box 9513, 2300 RA Leiden, The Netherlands}

\correspondingauthor{Anna-Christina Eilers}
\email{eilers@mit.edu} 

\begin{abstract}
We expect luminous ($M_{1450}\lesssim-26.5$) high-redshift quasars to trace the highest density peaks in the early universe. 
Here, we present observations of four $z\gtrsim6$ quasar fields using JWST/NIRCam in imaging and widefield slitless spectroscopy mode and report a wide range in the number of detected \oiii-emitting galaxies in the quasars' environments, ranging between a density enhancement of $\delta\approx65$ within a $2$~cMpc radius -- one of the largest proto-clusters during the Epoch of Reionization discovered to date -- to a density contrast consistent with zero, indicating the presence of a UV-luminous quasar in a region comparable to the average density of the universe. 
By measuring the two-point cross-correlation function of quasars and their surrounding galaxies, as well as the galaxy auto-correlation function, we infer a correlation length of quasars at $\langle z\rangle=6.25$ of $r_0^{\rm QQ}=22.0^{+3.0}_{-2.9}~{\rm cMpc}\,h^{-1}$, while we obtain a correlation length of the \oiii-emitting galaxies of $r_0^{\rm GG}=4.1\pm0.3~{\rm cMpc}\,h^{-1}$. By comparing the correlation functions to dark-matter-only simulations we estimate the minimum mass of the quasars' host dark matter halos to be $\log_{10}(M_{\rm halo, min}/M_\odot)=12.43^{+0.13}_{-0.15}$ (and $\log_{10}(M_{\rm halo, min}^{\rm [OIII]}/M_\odot) = 10.56^{+0.05}_{-0.03}$ for the \oiii-emitters), indicating that (a) luminous quasars do not necessarily reside within the most overdense regions in the early universe, and that (b) the UV-luminous duty cycle of quasar activity at these redshifts is $f_{\rm duty}\ll1$. Such short quasar activity timescales challenge our understanding of early supermassive black hole growth and provide evidence for highly dust-obscured growth phases or episodic, radiatively inefficient accretion rates. 
\end{abstract}

\keywords{dark ages, early universe --- quasars: supermassive black holes --- methods: data analysis} 

\section{Introduction}

The existence of supermassive black holes (SMBHs) with $\sim10^9-10^{10}\,M_\odot$ harbored in the center of luminous quasars at redshifts $z\gtrsim6$, when the universe was less than a billion years old, poses significant challenges to our understanding of black hole growth \citep[e.g.][]{Mazzucchelli2017, Banados2018, Yang2020, Wang2021, Yue2023, Bigwood2024}. In the standard picture of SMBH growth, the rest mass energy of the accreted matter is divided between radiation and black hole growth via the so-called radiative efficiency $\epsilon$, implying that the emission of quasar light is concurrent with the growth of the black hole. 
Every massive galaxy is thought to have undergone a luminous quasar phase, and thus high-redshift quasars represent the progenitors of the dormant SMBH population found in the centers of nearly all local bulge-dominated galaxies \citep{SmallBlandford1992, YuTremaine2002}. 
This fundamental link between high-redshift quasars and local galaxies is supported by the tight correlations between the masses of SMBHs and various galaxy properties, such as the stellar velocity dispersion or bulge mass \citep[e.g.][]{Magorrian1998, Gebhardt2000, FerrareseMerritt2000, HaeringRix2004}. 

In the current paradigm, black holes grow exponentially starting from an initial black hole seed $M_{\rm seed}$ during the lifetime $t_{\rm Q}$ of the quasar, which is the time that galaxies spent in the luminous quasar state, i.e.\ 
\begin{equation} 
M_{\rm BH}(t_{\rm Q}) = M_{\rm seed} \exp{(t_{\rm Q}/t_{\rm S})}. 
\end{equation}
The growth occurs on a characteristic timescale $t_{\rm S}$ known as the $e-$folding time, or ``Salpeter'' time \citep{Salpeter1964}, which is approximately
\begin{equation}
    t_{\rm S}\approx 0.45\,\left(\frac{\epsilon}{1-\epsilon}\right)\left(\frac{L}{L_{\rm Edd}}\right)^{-1}~\rm Gyr, 
\end{equation}
where $L$ and $L_{\rm Edd}$ denote the bolometric and Eddington luminosity of the quasar, respectively. Assuming a thin accretion disk around the black hole, the radiative efficiency $\epsilon$ ranges between $\sim6\%-34\%$, the exact value depending on the spin of the black hole \citep{ShakuraSunyaev1973, Thorne1974}.

Thus, 
long timescales of quasar activity -- comparable to the Hubble time $t_{\rm H}$ -- are required with nearly continuous Eddington-limited accretion, in order to explain the growth of the observed SMBHs at high redshift \citep[e.g.][]{TanakaHaiman2009, BegelmanVolonteri2017}. 
It follows that the fraction of cosmic time that galaxies shine as luminous quasars, known as the duty cycle $f_{\rm duty}=t_{\rm Q}/t_{\rm H}(z)$, is expected to be approximately unity for quasars in the early universe. 
This theoretical picture of SMBH formation predicts that early SMBHs form and grow in the centers of rare and massive dark matter halos, which constitute the highest density peaks of the cosmological density field \citep{EfstathiouRees1988, ColeKaiser1989, NusserSilk1993, Djorgovski2003, VolonteriRees2006, Sijacki2009, Costa2014, Trebitsch2019, Costa2023}. 
%
Thus, luminous quasars are expected to reside in massive overdensities at early cosmic times, which implies that one should find a high number of companion galaxies in their vicinity according to our $\Lambda$CDM cosmological model \citep[e.g.][]{MunozLoeb2008, Tinker2010}. The regions around some high-redshift quasars are predicted to constitute proto-clusters, progenitors of the galaxy clusters that populate the local universe, albeit the growth of these structures across cosmic time sensitively depends on a variety of environmental factors \citep[e.g.][]{Angulo2012}. 

However, to date the search for the expected abundance of companion galaxies around $z\gtrsim6$ quasars has led to inconclusive results. While some studies find strong galaxy enhancements in quasar fields using mostly narrow-band filters to capture the \lya\ emission of galaxies in the quasars' environments \citep[e.g.][]{Kim2009, Morselli2014}, others do not \citep[e.g.][]{Banados2013, Simpson2014, Mazzucchelli2017_clustering}. The first, recently published observations with JWST/NIRCam WFSS of two high-redshift quasar fields revealed an abundance of \oiii-emitting galaxies in the quasars' environment, indicating that these particular quasars lie indeed in highly overdense regions in the universe \citep{Wang2023, Kashino2023}. It is unclear whether the conflicting results arise from the limited sensitivity, velocity offsets between the narrow-band filter and the quasars' redshifts, or a too small field of view of some of the observations \citep[e.g.][]{Overzier2016}, or alternatively, at least some quasars may inhabit less massive hosts at $z\gtrsim 6$ than expected. 

The mass of the dark matter halos hosting these luminous quasars can be inferred by means of their clustering properties. The two-point correlation function is a powerful tool to probe the connection between the observed light and host dark matter halos \citep[e.g.][]{Osmer1981, ShanksBoyle1994, Croom2001, Croom2005, Shen2007}. 
%
At low and intermediate redshifts of $1\lesssim z\lesssim 4$ the two-point auto-correlation function has been studied using several thousands of quasars distributed over large areas on the sky. Across this wide redshift range most quasar clustering studies find similar masses of the dark matter halos that host luminous quasars of a few times $10^{12}\,M_\odot$ \citep[e.g.][]{Porciani2004, Croom2005, Shen2007, Myers2007, Coil2007, Ross2009, Padmanabhan2009, TrainorSteidel2013, Eftekharzadeh2015, He2018}. This implies that quasars at intermediate redshifts are highly biased tracers of the underlying density field and reside within some of the most massive dark matter halos at these cosmic times. At $z\sim1$, however, the dark matter halo mass is close to the characteristic mass in the Press-Schechter mass function \citep{PressSchechter1974}, indicating that quasars in the local universe are largely unbiased \citep{Croom2005}. Thus, we observe an overall increase in the quasars' bias factor with redshift, i.e.\ the ratio of the host halo mass for a typical quasar to the mean halo mass at the same epoch rises with increasing redshift. 

In turn, this implies an increase in the quasars' duty cycle from $\lesssim1\%$ at low redshifts, i.e.\ $z\lesssim1$ \citep{Croom2005, Padmanabhan2009, Laurent2017}, to a few tens of percent at $z\sim 4$ (\citet{Shen2007, Pizzati2023}, but see \citet{Eftekharzadeh2015} who report a lower duty cycle at similar redshifts). Thus, the quasar activity timescales at these redshifts are expected to be relatively long, i.e.\ $t_{\rm Q}>10^8$~years, therefore providing sufficient time to grow the observed SMBHs. 





At redshifts $z>6$, however, this approach to determine the host dark matter halo masses by measuring the quasars' auto-correlation function becomes unfeasible as the space density of luminous ($M_{1450}<-27$) quasars drops dramatically to less than $1\,\rm Gpc^{-3}$ \citep{Wang2019, Schindler2023}. Using a much fainter ($M_{1450}\gtrsim-25$), but more numerous population of quasars a first attempt of the auto-correlation measurement at $z\sim 6$ has recently been reported by the Subaru High-z Exploration of Low-luminosity Quasars (SHELLQs) Collaboration \citep{Arita2023}, indicating host dark matter halo masses similar or slightly higher than those found at lower redshifts, but uncertainties are large. However, while auto-correlation measurements of quasars at these high redshifts remain challenging, the complementary quasar-galaxy cross-correlation measurement is still viable. Assuming that both quasars and galaxies trace the same underlying dark matter density distribution, their respective auto-correlation functions determine the cross-correlation function between these two classes of objects.  

During the last year, JWST has revolutionized our ability to study the early universe, enabling us to reveal and spectroscopically confirm faint galaxies down to $m_{\rm AB}\sim29\rm\,mag$ within the environment of high-redshift quasars. Using NIRCam in imaging and WFSS mode, JWST recently obtained large mosaic observations in five quasar fields at $6\lesssim z\lesssim7$ as part of the \textit{Emission line galaxies and Intergalactic Gas in the Epoch of Reionization} (EIGER) Collaboration \citep{Kashino2023}. The filter for the grism spectra was tuned to spectroscopically identify galaxies in the quasar fields by means of their \oiii\ emission line doublet $\lambda\lambda4960,5008$. 
Using this data set, we present in this paper for the first time the quasar-galaxy cross-correlation measurement at $z\gtrsim6$. Combined with the auto-correlation of the \oiii-emitting galaxies, we infer the clustering properties of these high-redshift luminous quasars, and determine their host dark matter halo mass as well as their duty cycle.

In \S~\ref{sec:data} we will first describe the JWST/NIRCam observations of the high-redshift quasar fields analyzed in this study. We will briefly comment in \S~\ref{sec:over} on the large diversity in the observed abundance of galaxies that we observe in the environments of the quasars, before measuring the two-point cross-correlation function of the quasars and their surrounding galaxies, as well as the galaxy-galaxy auto-correlation function in \S~\ref{sec:clustering}. In \S~\ref{sec:DM} we infer the characteristic mass of the quasars' host dark matter halos based on their clustering properties, and estimate their duty cycle in \S~\ref{sec:duty_cycle}. We discuss the implications of our results in \S~\ref{sec:discussion} before summarizing in \S~\ref{sec:summary}. 

Throughout this work we adapt a flat $\Lambda$CDM cosmology with $H_0 = 67.66\,\rm km\,s^{-1}\,Mpc^{-1}$, $\Omega_M = 0.3097$, and $\Omega_\Lambda = 0.6889$ \citep{Planck2018}.

\section{JWST/NIRCam Observations of High-Redshift quasar fields}\label{sec:data}

\begin{figure*}[!t]
    \centering
    \includegraphics[width=\textwidth]{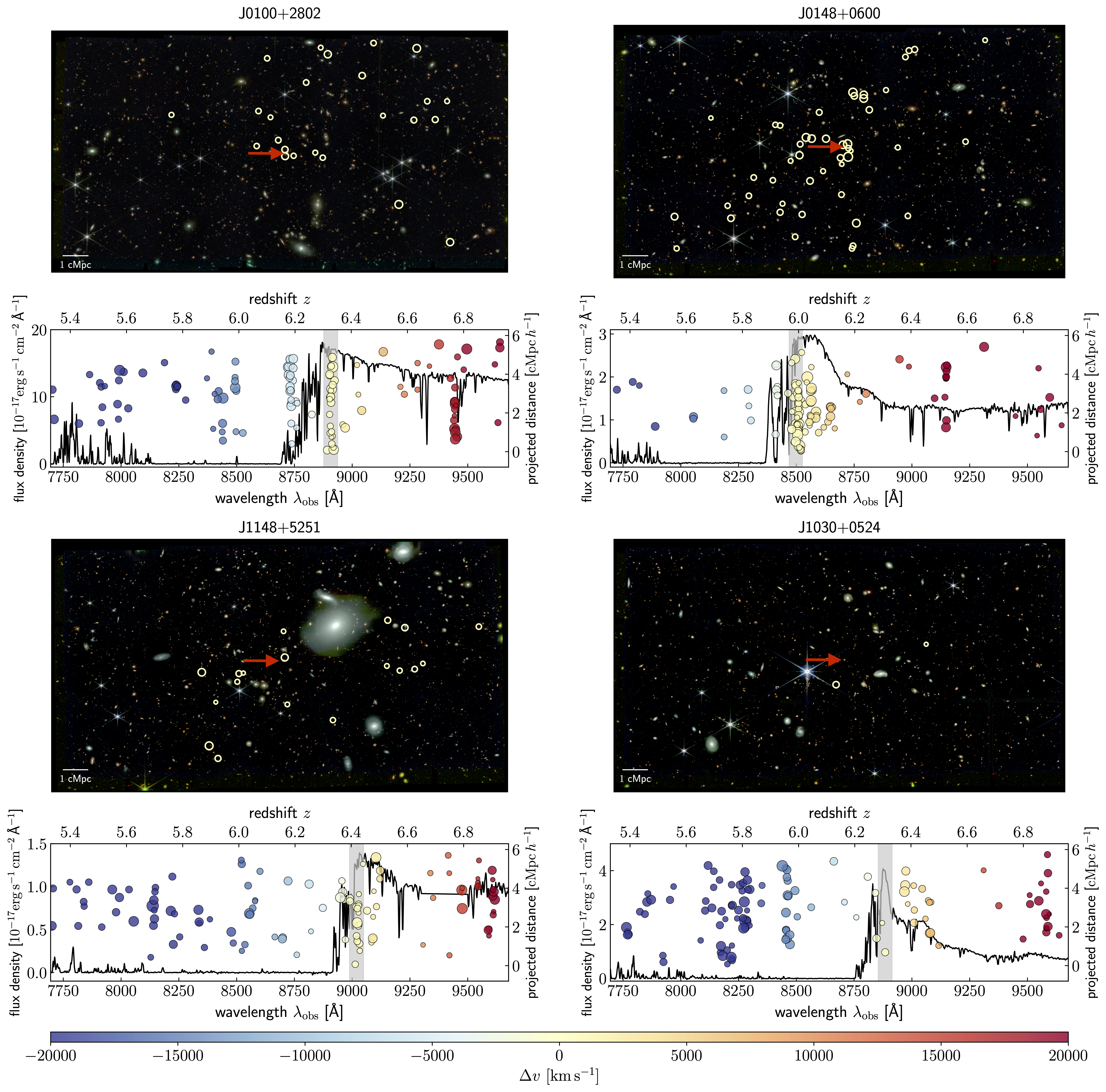}
    \caption{\textbf{Observations of four $z\gtrsim 6$ quasar fields.} Top panels show the RGB images constructed from NIRCam imaging data (F115W, F200W and F356W) of each quasar field. Bottom panels show the optical and NIR spectra of the quasars observed with X-Shooter/VLT and FIRE/Magellan (J0100+2802, J0148+0600, J1030+0524), or MOSFIRE+ESI/Keck (J1148+5251). The colored data points show the redshifts and projected distances of all \oiii-emitting galaxies detected in the NIRCam grism spectra above a luminosity of $L_{\rm[O\,III], 5008}=10^{42}\,\rm erg\,s^{-1}$; their color corresponds to their velocity offset with respect to the quasars' redshifts and their size correlates with the logarithm of their luminosity, $\log_{10}L_{\rm[O\,III], 5008}$. The grey shaded area denotes the velocity interval of $\left|\Delta v\right| \leq v_{\rm max}$ around the redshift of the quasar. All galaxies that fall within this velocity range are also circled in the respective RGB images to show the spatial distribution around the quasar (indicated by a red arrow). }
    \label{fig:rgb}
\end{figure*}

The EIGER Collaboration is observing a total of six $z\gtrsim 6$ quasar fields utilizing NIRCam in WFSS and imaging mode (ID: $\#1243$). Of these six fields, the observations of five quasar fields were completed at the time of writing, and their analysis is presented in this work. Four out of five fields are used to study the quasar-galaxy clustering (as in one field our setup does not cover the \oiii-emission at the quasar's redshift), while all five quasar fields are used when measuring the galaxy auto-correlation (see \S~\ref{sec:clustering}). 

\subsection{Data Reduction}

An overview of the observing program as well as a detailed description of the data reduction can be found in \citet{Kashino2023} and \citet{Matthee2023}. We will briefly summarize the setup of the program and the main steps of the data reduction here. 

The NIRCam imaging and WFSS observations of the five quasar fields were conducted between August 22, 2022 and June 2, 2023. The observations consist of four visits per quasar field forming a $2\times2$ overlapping mosaic pattern, which provides an area of approximately $3'\times 6'$ per field centered on the quasar. The central $40''\times 40''$ region around the quasar is covered by all four mosaic tiles. 

The WFSS observations use the grism $R$ in the long wavelengths (LW) channel combined with the F356W filter. This yields spectra covering the wavelengths between $3.1–4.0\,\mu$m dispersed along the detector rows with a spectral resolution of $R\approx 1500$. 
The two NIRCam modules A and B disperse the spectra in opposite directions. These reversed grism spectra facilitate significantly the identification of the source objects in the direct images, and enable us to remove contaminating lines from other sources. 

Our observations employ a three-point primary dither using \texttt{INTRAMODULEX} and a four-point subpixel dither, which yields a total of $12$ exposures per visit and module for the direct images in the two short wavelength (SW) filters (F115W and F200W), and $24$ WFSS images. By co-adding the multiple exposures, the total exposure time in a single visit amounts to $4380\,\rm s$ per filter for the SW imaging and $8760\,\rm s$ for WFSS. We obtained additional direct imaging in the LW channel F356W filter after the WFSS observations, with F200W in the SW channel. Given the small-scale dithering and larger-scale mosaic pattern, the total exposure time varies across the field, ranging from $1.5\,\rm ks$ at the edges sampled by only a single primary dither up to $13\,\rm ks$ in the center of each field for SW imaging, and from $2.9$~ks up to $35$~ks for WFSS \citep{Kashino2023}. 

The WFSS data were reduced with a combination of the \texttt{jwst} pipeline (version 1.9.4) and additional custom processing steps. We first process the raw exposure files with the \texttt{Detector1} step from the pipeline and use the \texttt{Spec2} step to assign the astrometric solution for each image. The data is flat-fielded by applying the \texttt{Image2} step and sky background variations as well as $1/f$ noise are removed using the median counts in each column. We construct an emission line image by applying a running median filter along each row to remove any continuum emission \citep[see][for details]{Kashino2023}. 


\begin{deluxetable*}{cccCCCCCCC}[!t]
\setlength{\tabcolsep}{2pt}
\renewcommand{\arraystretch}{1.1}
\tablecaption{Overview of the number of \oiii-emitting galaxies in the quasar fields. \label{tab:observations}}
\tablehead{\colhead{quasar field} & \colhead{RA}& \colhead{DEC}&\dcolhead{z} & \dcolhead{M_{1450}}&\dcolhead{\log_{10}(M_{\rm BH}/M_\odot)} &\dcolhead{N_{\rm tot}} & \dcolhead{N_{\rm gal}} & 
\dcolhead{V}&\dcolhead{\delta} \\ 
 & [hms]& [dms]& & & & (\geq10^{42}\,\rm erg\,s^{-1})&(|\Delta v|\leq v_{\rm max})& [\rm cMpc^{3}]& (r<2\,{\rm cMpc}) }
\startdata
\rm J0148+0600 & 01:48:37.64& $+$06:00:20.06&5.99^a &-26.99& 9.89^{+0.05}_{-0.06}&113 & 47 & 2469 & 65\pm15 \\ 
\rm J1030+0524 & 10:30:27.10&	$+$05:24:55.00&6.308^b &-27.39& 9.19\pm 0.01&112 & 2&  2631&3\pm4\\ 
\rm J0100+2802 &01:00:13.02&	$+$28:02:25.84&6.3270^a&-27.62& 10.06\pm0.01 &121 & 24 & 2554&29\pm10\\ 
\rm J1148+5251 & 11:48:16.64 & $+$52:51:50.30&6.4189^a & -29.14& 9.64\pm 0.01&110 & 19 & 2650 &16\pm8\\ 
\rm J1120+0641 &11:20:01.48&$+$06:41:24.30&7.0848^a& -26.63& 9.08\pm 0.03&66 & - & - & -\\
\enddata
\tablecomments{The columns denote the quasar field, the quasars' coordinates RA and DEC, the quasars' redshifts (determined by means of $a$: \cii\ or $b$: \mgii\ emission lines) and absolute magnitude at $1450$~{\AA} in the rest-frame, the masses of the quasars' SMBHs determined via the H$\beta$ emission line \citep{Yue2023} without any systematic uncertainties of $\sim 0.5$~dex, the number of \oiii-emitting galaxy systems above a luminosity limit of $L_{\rm[O\,III], 5008}=10^{42}\,\rm erg\,s^{-1}$ detected in the whole quasar field, the number of these galaxies within $|\Delta v|\leq1000\,\rm km\,s^{-1}$ of the quasar's redshift, the total observed survey volume $V$ around the quasar within $\pm1000\,\rm km\,s^{-1}$, and the observed overdensity $\delta$ within a radius of $2~\rm cMpc$ around the quasar. }
\end{deluxetable*}

\subsection{Identification of \oiii-Emitting Galaxies}\label{sec:ID}

The spectral range covered by the filter F356W of $3.1\leq\lambda\leq4.0\,\mu$m enables us to identify \oiii-emitters within the redshift range $5.33\lesssim z\lesssim 6.96$. This implies that we can study the \oiii-emitting galaxies in the environment of four of the targeted quasars, while for one quasar, J1120+0641 at $z=7.0848$, the \oiii\ emission lines lie outside of the observed redshift range. The galaxies observed in this field, however, will be included in the measurement of the galaxy-galaxy auto-correlation function (\S~\ref{sec:auto-corr}). 

We employ two complementary methods to identify \oiii-emitting galaxies from the emission line images. The first approach starts with the co-added emission line image and searches for combinations of emission lines that could plausibly arise from doublets (\oiii\ $\lambda\lambda\,4960, 5008$) or triplets (\oiii\ $\lambda\lambda\,4960, 5008$ + H$\beta$ $\lambda4861$) using \texttt{SExtractor}. To evaluate the validity of these detections, we then try to identify a source galaxy in the F356W direct image using the grism model, which enables us to estimate the location on the sky for a given position in the grism image. 


As a second method to identify \oiii-emitters, we employ an inverse approach and use the catalog of sources identified in the F356W images as a starting point. For each source we extract a 2D spectrum from the median-filtered grism images and run \texttt{SExtractor} to identify emission lines. \oiii-emitters are identified if emission line doublets or triplets are detected at $\geq3\sigma$ close to the expected locus of the spectral trace of the object. 
The list of detected \oiii-emitters from both approaches are afterwards combined and cross-validated. We estimate the systematic uncertainty on the derived redshifts to be about $100\,\rm km\,s^{–1}$ due to the inaccuracy of the grism model \citep{Kashino2023}.

Note that our identification algorithm for \oiii-emitters has been fine-tuned and improved since the publication of the first results from the EIGER Collaboration \citep{Kashino2023, Matthee2023}. This results in an increased number of discovered \oiii-emitting galaxies. Within the field around quasar J0100+2802 we now discover a total of $180$ \oiii-emitting galaxy systems ($213$ individual galaxies, see \S~\ref{sec:grouping} for details on how we define a group) compared to the previously reported $117$. Our total sample of individual \oiii-emitters in all five quasar fields amounts to $861$ (Kashino et al.\ in prep.). 

\subsection{Groups of \oiii-Emitting Galaxies}\label{sec:grouping}

When analyzing the distribution of \oiii-emitting galaxy pairs as a function of their angular separation, we find a significant excess in the number of ``clumps'', i.e.\ pairs or multiples, on very small scales at separations $\lesssim2''$, which corresponds to approximately $12~\rm kpc$ at $z=6$ \citep[see Fig. $4$ in][]{Matthee2023}. All of these \oiii-emitting systems within $2''$ separation are similarly closely separated in redshift ($|\Delta v|\lesssim 600\,\rm km\,s^{-1}$), suggesting that these components are physically associated with each other. Consequently, we merge all \oiii-emitters within an angular separation $<2''$ and $|\Delta v|< 1000\,\rm km\,s^{-1}$ together into a single system. The motivation for this is to mitigate uncertainties in the identification of individual objects that can arise due to the choice of deblending parameters used in \texttt{SExtractor}. 

This merges about $\sim 13\%$ of the detected indivdiual \oiii-emitters and results in a total of $750$ \oiii-emitting \textit{systems} discovered in the five quasar fields. Since the completeness of our detections drops significantly below a luminosity limit of $L_{\rm[O\,III], 5008}<10^{42}\,\rm erg\,s^{-1}$ (see \S~\ref{sec:completness}), we restrict ourselves in this study to the $533$ systems more luminous than this threshold (see Tab.~\ref{tab:observations}). 
Furthermore, we restrict our analysis to systems that have been found using the second method to identify \oiii-emitters starting from the source catalog identified in the F356W image (see \S~\ref{sec:ID}), since the completeness of our observations is assessed using this method (see \S~\ref{sec:completness}). This step only removes $11$ objects, with a total of $522$ systems remaining for our analysis. 

Fig.~\ref{fig:rgb} shows our observations for four of the quasar fields, for which the quasars' environment can be studied with our setup. For each field, we show in the top panels the NIRCam images (F115W, F200W and F356W) indicating the location of the quasar in the center as well as all \oiii-emitting galaxies within the quasar's environment, i.e. within $|\Delta v|\leq 1000\,\rm km\,s^{-1}$ 
from the quasar's systemic redshift. 
The bottom panels show the quasar spectra observed with the ground-based spectrographs FIRE \citep{Simcoe2008} on the Magellan Telescopes, X-Shooter \citep{Vernet2011} on the Very Large Telescope and MOSFIRE \citep{MOSFIRE} and ESI \citep{ESI} on the Keck Telescopes \citep[see][for details on the ground-based quasar spectra]{Eilers2023, Durovcikova2024}, as well as all detected \oiii-emitting systems in the quasar field as a function of redshift and projected distance from the quasar. 


\subsection{Modeling the completeness of our observations and creating random galaxy catalogs}\label{sec:completness}

In order to measure the clustering in the quasar fields as well as the excess in galaxies in the quasars' environments compared to a blank field, we need to construct a set of ``random'' galaxy catalogs. These catalogs contain a large number of sources with similar properties as the observed galaxies, but randomly distributed across the survey volume. 

Our selection function is not trivial due to the 4-point mosaic, the different sensitivities of NIRCam's module A and B, as well as the wavelength dependence of the noise over our field of view. 
To this end, we inject random sources in our survey volume and forward model their spectra through our survey coverage and completeness. 
Their \oiii-luminosities are randomly drawn from the \oiii-luminosity distribution function \citep[][with updated model parameters from Mackenzie et al.\ in prep.]{Matthee2023} and a minimum luminosity of $L_{\rm[O\,III], 5008}=10^{42}\,\rm erg\,s^{-1}$ in the \oiii-emission line at $5008$~{\AA}. Note that the \oiii-luminosity function is estimated over the entire redshift range, but masking all galaxies at the quasars' redshifts, such that no biases due to potential overdensities in the quasars' environments arise. The spatial positions and redshifts are randomly chosen within the observed survey volume. We then check the spatial and spectral coverage of this source in our survey, and ensure that both emission lines of the \oiii-doublet, i.e.\ at $4960$~{\AA} and $5008$~{\AA}, would have been observed in our setup. Since both our methods to identify \oiii-emitting galaxies are based on the observability of the line doublet (see \S~\ref{sec:ID}), we would miss all sources in our catalog for which only a single line would be observed. 


To model the completeness of our line-selected sample into the random catalogs, we apply the completeness models that will be described in a forthcoming companion paper (Mackenzie et al.\ in prep.). In short, each field has an individual model, which consists of a noise cube and a completeness interpolating function. The noise cube represents the typical propagated error in a 2D stacked spectrum as a function of position and wavelength. The function is constructed by injecting millions of mock sources into the 2D spectra and re-running the \oiii-emitter identification step. Thus, the measured completeness is not a step function at some limiting flux, but instead is a smooth function which we interpolate from the injection experiments. This function then translates the flux of the weaker of the two \oiii\ emission lines, i.e. $4960$~{\AA}, divided by the local error, to a completeness fraction. Dividing by the error essentially re-scales the flux such that the completeness function is independent of the local sensitivity or redshift. To create the random galaxy catalogs we calculate these completeness maps of each EIGER field, and then draw random samples using the completeness as a probability threshold. 
Since our observations become highly incomplete for sources below a luminosity of $L_{\rm[O\,III], 5008}= 10^{42}\,\rm erg\,s^{-1}$, we only consider galaxies above this luminosity limit for our analysis, which corresponds to a line flux of for the weaker \oiii-emission line of $f_{\rm[O\,III], 4960}= 0.7\times10^{-18}\,\rm erg\,s^{-1}\,cm^{-2}$ at $z\approx6.3$. 

We estimate an average completeness of $\sim 41\%$ for galaxies with a luminosity of $L_{\rm[O\,III], 5008}= 10^{42}\ \rm erg\,s^{-1}$ at $6.0<z<6.4$, which increases to $\sim 98\%$ for $L_{\rm[O\,III], 5008}= 10^{43}\,\rm erg\,s^{-1}$. Note that these completeness estimates are averaged over the five quasar fields and the full spatial coverage of each mosaic, for regions closer to the quasar the completeness is higher. For more details on the completeness evaluation of our observations we refer the reader to \citep[][Mackenzie et al.\ in prep.]{Matthee2023}. 

\begin{deluxetable*}{CCCCCCCCC}[!t]
\setlength{\tabcolsep}{8pt}
\renewcommand{\arraystretch}{1.1}
\tablecaption{Quasar-galaxy cross-correlation and galaxy-galaxy auto-correlation measurements at $\langle z\rangle=6.25$. \label{tab:xcorr}}
\tablehead{\dcolhead{R_{\rm min}} & \dcolhead{R_{\rm max}} &\dcolhead{\langle \rm QG\rangle} &  \dcolhead{\langle \rm QR\rangle}&\dcolhead{\langle \rm GG\rangle} &\dcolhead{\langle \rm GR\rangle}&\dcolhead{\langle \rm RR\rangle}& \dcolhead{\chi_{\rm QG}}& \dcolhead{\chi_{\rm GG}}\\ 
\dcolhead{[{\rm cMpc}\,h^{-1}]} & \dcolhead{[{\rm cMpc}\,h^{-1}]} }
\startdata
0.06 & 0.10& 2 & 3.3\times10^{-3}& 3& 0.10 & 0.11 & 604.2\pm 427.9 & 26.0\pm17.2\\
0.10 & 0.18 & 3 & 1.4\times10^{-2} & 6 & 0.33 & 0.34  & 210.8\pm122.3 & 15.8\pm8.0 \\
0.18 & 0.31 & 3 & 4.3\times10^{-2} & 16 & 1.02 & 1.05 & 68.3\pm 40.0 & 13.2\pm4.2\\
0.31 & 0.56 & 4 & 0.14 & 29 & 3.16 & 3.24 & 27.0\pm14.0 & 7.0\pm 2.0\\
0.56 & 1.00 & 7 & 0.44 & 66 & 9.48 & 9.65 & 15.0\pm6.1 & 4.9\pm1.1\\
1.00 & 1.77 & 21 & 1.25 & 117 & 26.92 & 27.29 & 15.8\pm3.7 & 2.3\pm0.5\\
1.77 & 3.15 & 30 & 3.69 & 203 & 67.07 & 68.39 & 7.1\pm1.5 & 1.0\pm0.3 \\
3.15 & 5.60 & 21 & 4.55 & 260 & 113.43 & 116.50 & 3.6\pm1.0 & 0.3\pm0.2
 \enddata
\end{deluxetable*}

\section{Large Diversity in Quasar Environments}\label{sec:over}

\begin{figure}[!t]
    \centering
    \includegraphics[width=0.48\textwidth]{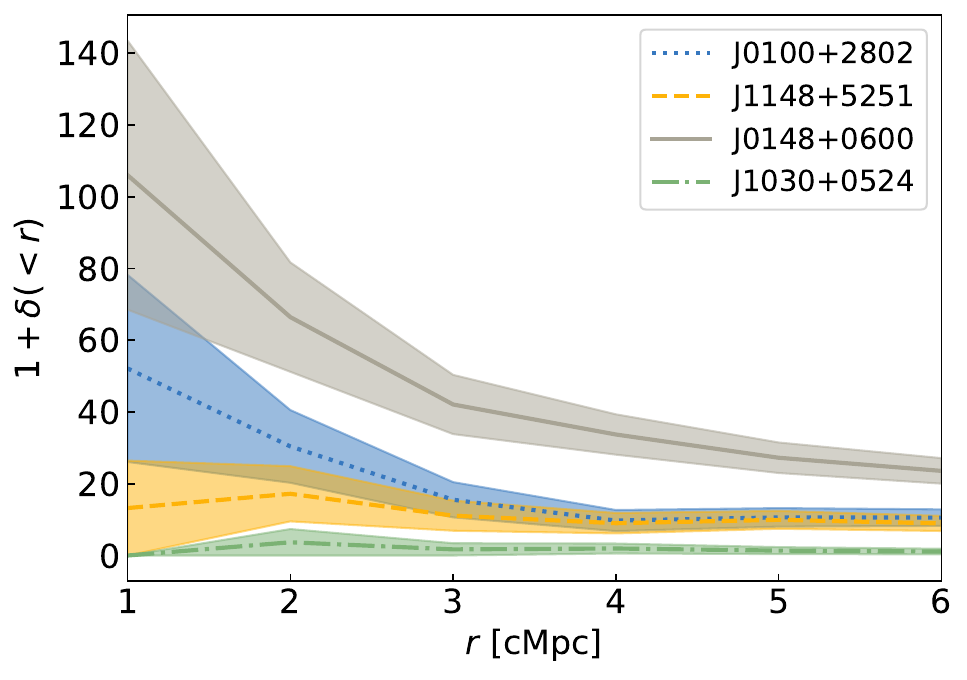}
    \caption{Density contrast of \oiii-emitters around the four $z\gtrsim 6$ quasars analyzed in this work. The overdensity is measured in cylinders around the quasars with radius $r$ and a depth of $|\Delta v| = 1000\,\rm km\,s^{-1}$ around the quasars' redshifts. }
    \label{fig:over}\label{fig:delta}
\end{figure}

Current SMBH formation models predict that luminous quasars in the high-redshift universe trace the highest-density peaks and thus reside in a significantly overdense environment \citep[e.g.][]{Sijacki2009}. 
To this end we estimate the density contrast, i.e.\ $\delta=N_{\rm gal}/N_{\rm exp}-1$, by comparing the number of galaxies detected in close vicinity to the quasars $N_{\rm gal}$, to the number of galaxies one would expect within the same observed volume $N_{\rm exp}$ in a blank ``random'' part of the universe. In this case, $N_{\rm gal}$ denotes simply the number counts of \oiii-emitting galaxy systems with $|\Delta v|\leq1000\,\rm km\,s^{-1}$ from the quasars' systemic redshift and within different radii. The expected number of objects $N_{\rm exp}$ is estimated by multiplying the galaxy density $n_{\rm gal}$ inferred from the \oiii-luminosity function \citep[][Mackenzie et al.\ in prep.]{Matthee2023} by the completeness map derived in \S~\ref{sec:completness} and by the observed volume $V$ around the quasar, i.e.\ a cylinder with a given radius and depth of $\pm 1000\,\rm km\,s^{-1}$. 
We estimate the uncertainties on the density contrast from Poisson noise on the galaxy counts. 


Our results are shown as a function of radius $r$ for the four quasar fields in Fig.~\ref{fig:over} and suggest a strong diversity among the observed quasar fields. We find significant overdensities around three of the four observed quasars. In close vicinity of the quasars within $r<2$~cMpc we find density enhancements of $\delta(r<2\,{\rm cMpc})=29\pm10$ and $\delta(r<2\,{\rm cMpc})=16\pm8$ around the quasars J0100+2802 and J1148+5251, respectively. With a total of $N_{\rm gal}=47$ galaxies observed in the complete observed volume around the quasar J0148+0600, this quasar resides in a spectacular proto-cluster within the Epoch of Reionization with an overdensity of $\delta(r<2\,{\rm cMpc})=65\pm15$. 

Interestingly, one of the targeted quasar fields around J1030+0542, shows only $2$ \oiii-emitting galaxies within the $|\Delta v|\leq1000\,\rm km\,s^{-1}$ velocity interval from the quasar and therefore no galaxy enhancement compared to a blank field, i.e.\ $\delta(r<2\,{\rm cMpc})=3\pm4$. We will discuss the possible implications of the highly diverse quasar environments in \S~\ref{sec:discussion}. All measurements and galaxy counts across the whole  observed volume, as well as the density enhancements at $r<2$~cMpc, are listed in Tab.~\ref{tab:observations}. 

\section{The Clustering Properties of High-redshift Quasars}\label{sec:clustering}


In this section we present measurements of the cross-correlation function between the quasars and the \oiii-emitting galaxies in their environment (\S~\ref{sec:xcorr}), as well as of the auto-correlation function of  \oiii-emitters outside of the quasars' environments (\S~\ref{sec:auto-corr}). By jointly fitting these two correlation functions we will infer the clustering properties of the luminous quasars themselves (\S~\ref{sec:fit_QG} and \S~\ref{sec:fit}). 

\subsection{The Quasar-Galaxy Cross-Correlation Function}\label{sec:xcorr}

We first measure the volume-averaged projected cross-correlation function between the quasars and the surrounding \oiii-emitting galaxies, i.e.\ $\chi_{\rm QG}(R_{\rm min}, R_{\rm max})$, which is a dimensionless quantity defined as the real-space quasar-galaxy cross-correlation function $\xi_{\rm QG}(R, Z)$ integrated over and normalized by a comoving volume $V$ \citep[e.g.][]{Hennawi2006_clustering, GarciaVergara2017}, i.e.\  
\begin{align}
    \chi_{\rm QG}&(R_{\rm min}, R_{\rm max}) = \nonumber \\ 
    &\frac{2}{V}\int_{R_{\rm min}}^{R_{\rm max}}\int_{0}^{Z_{\rm max}}\xi_{\rm QG}(R, Z)\,2\pi R\,{\rm d}R\,{\rm d}Z. \label{eq:chi}
\end{align}
We consider galaxies with a maximum velocity difference from the central quasar of $\left|\Delta v\right| \leq v_{\rm max}$ with $v_{\rm max}=1000\,\rm km\,s^{-1}$, and thus $Z_{\rm max}=v_{\rm max}(1+z)/H(z)$, where $H(z)$ denotes the Hubble parameter at redshift $z$. The integrated volume $V$ is a cylindrical shell with inner and outer radii $R_{\rm min}$ and $R_{\rm max}$, respectively, as well as a depth of $Z_{\rm max}$. We integrate over the redshift range, in order to avoid the need to model redshift-space distortions and redshift uncertainties. 

Following previous work \citep[e.g.][]{GarciaVergara2017} we choose the estimator 
\begin{equation}
    \chi_{\rm QG}(R_{\rm min}, R_{\rm max}) = \frac{\langle \rm QG\rangle}{\langle \rm QR\rangle}-1, 
\end{equation}
where $\langle \rm QG\rangle$ denotes the number of \textit{detected} galaxies in radial bins around the quasar, while $\langle \rm QR\rangle$ describes the number of \textit{expected} galaxies within the same volume in a random ``blank'' field. Thus, in order to infer $\langle \rm QR\rangle$, we calculate the number of pairs between the quasar and galaxies from our random catalog within the same radial bins and $\left|\Delta v\right| \leq v_{\rm max}$, and normalize them to the number of galaxies we would have expected to find within the same observed cylindrical volume in a ``blank'' field. As before, we estimate the expected quasar-galaxy pair counts by means of the number density of galaxies inferred from the \oiii-luminosity function multiplied by the observed volume in each shell and the completeness of our survey volume. Uncertainties on the cross-correlation measurements are estimated using Poisson noise on the detected galaxy counts\footnote{Note that in reality the uncertainties on the correlation measurements arise from a combination of Poisson counting errors as well as cosmic variance. The latter errors are, however, challenging to estimate and require extensive cosmological simulations, and thus we approximate the uncertainties by the Poisson errors alone.}. 
Our measurements for $\chi_{\rm QG}$ are listed in Tab.~\ref{tab:xcorr} and shown in  Fig.~\ref{fig:acf_ccf}. 


\subsection{Galaxy-Galaxy Auto-Correlation Function}\label{sec:auto-corr}

Analogous to Eqn.~\ref{eq:chi}, we calculate the volume-averaged projected auto-correlation function of galaxies $\chi_{\rm GG}(R_{\rm min}, R_{\rm max})$ by integrating the real space galaxy-galaxy auto-correlation function $\xi_{\rm GG}(R, Z)$ in shells of volume $V$. We consider all galaxies within the redshift interval $5.95\leq z\leq 6.55$, which is approximately centered around the mean redshift of the quasars, $\langle z\rangle\approx6.25$. Again, galaxies within the quasars' environments are excluded from the auto-correlation measurement. We then calculate pairs of galaxies separated by $\left|\Delta v\right| \leq v_{\rm max}$. 

\begin{figure}[!t]
    \centering
    \includegraphics[width=0.47\textwidth]{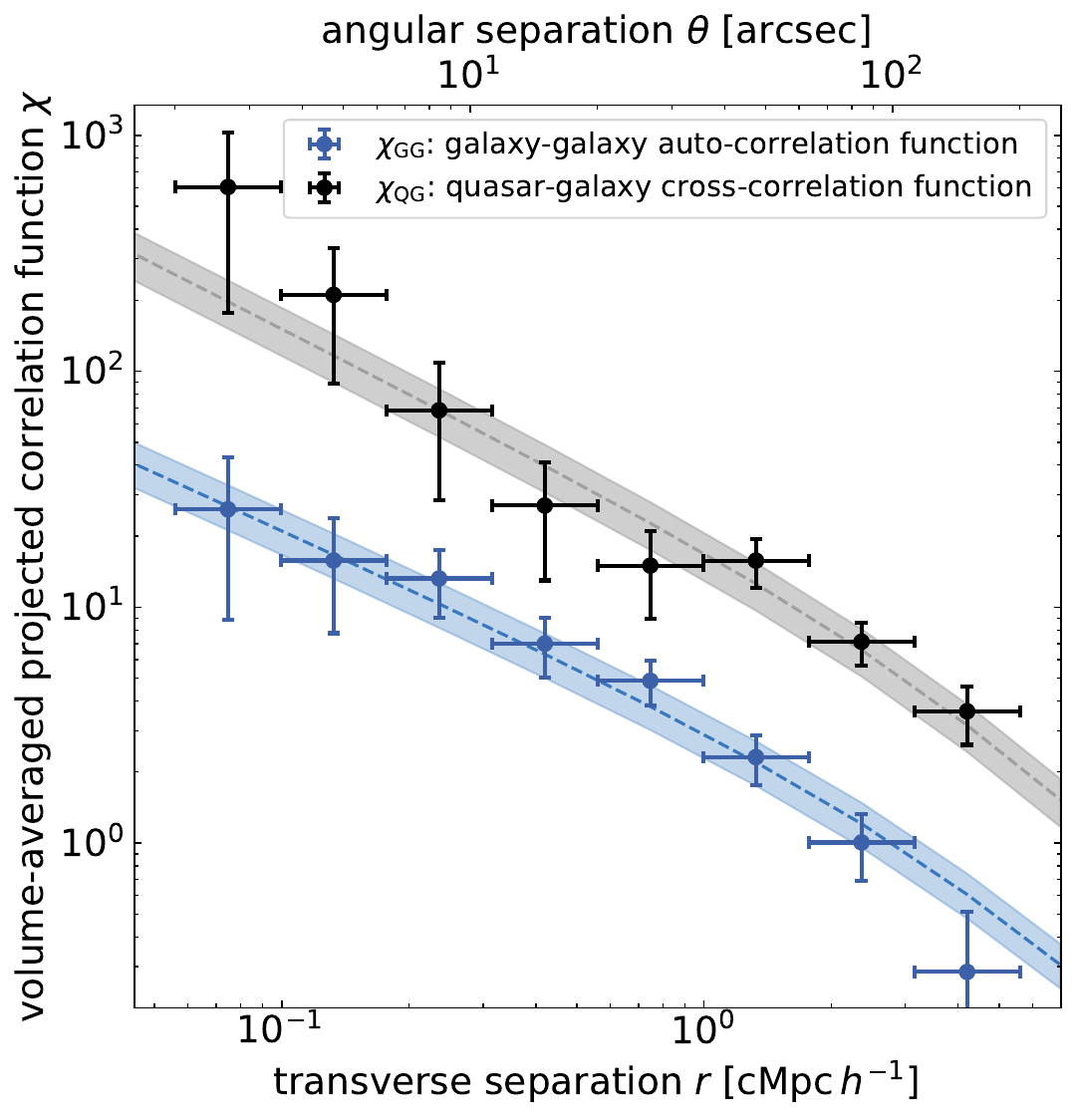}
    \caption{The volume-averaged projected quasar-galaxy cross-correlation function is shown as black data points, while the blue data points show the galaxy-galaxy auto-correlation function at an average redshift of $\langle z\rangle=6.25$. The dashed lines and shaded areas show the best power-law fits with $2\sigma$ uncertainties when jointly fitting both sets of measurements while keeping the slopes $\gamma_{\rm GG}=1.8$ and $\gamma_{\rm QG}=2.0$ fixed (see \S~\ref{sec:fit}). }
    \label{fig:acf_ccf}
\end{figure}

We choose the Landy-Szalay estimator \citep{LandySzalay1993}, i.e. 
\begin{equation}
    \chi_{\rm GG}(R_{\rm min}, R_{\rm max}) = \frac{\langle \rm GG\rangle - 2 \langle \rm GR\rangle + \langle \rm RR\rangle}{\langle \rm RR\rangle}-1, 
\end{equation}
where $\langle \rm GG\rangle$, $\langle \rm GR\rangle$ and $\langle \rm RR\rangle$ denote the number of galaxy pairs, pairs of detected galaxies and galaxies in the random catalog, as well as pairs of galaxies within the random catalog, respectively. The terms $\langle \rm GR\rangle$ and $\langle \rm RR\rangle$ are normalized such that the number of random galaxies would match the number of detected galaxies in each quasar field within the considered redshift interval. 
Uncertainties on the measurements are approximated by the Poisson noise on the number of detected galaxies. 
%
The measurements of the volume-averaged projected galaxy-galaxy auto-correlation function are shown in Fig.~\ref{fig:acf_ccf} and listed in Tab.~\ref{tab:xcorr}. 


\subsection{Fitting the Cross- and Auto-Correlation Functions}\label{sec:fit_QG}

The volume-averaged projected correlation functions $\chi_{\rm QG}$ and $\chi_{\rm GG}$ represent integrals of the real-space correlation functions $\xi_{\rm QG}$ and $\xi_{\rm GG}$ as shown in Eqn.~\ref{eq:chi}. We parameterize the real-space cross- and auto-correlation functions by a power-law, i.e.\
\begin{equation}
\xi(r)=\left(r/r_0\right)^{-\gamma} \label{eq:pl}
\end{equation}
where $r=\sqrt{R^2+Z^2}$. Thus, the real-space correlation functions are governed each by two parameters, i.e.\ their slopes $\gamma$, denoted as $\gamma_{\rm QG}$ and $\gamma_{\rm GG}$ for the cross- and auto-correlation, respectively, and their scale lengths $r_0$, denoted as $r_0^{\rm QG}$ and $r_0^{\rm QQ}$. 

We use a Markov Chain Monte Carlo algorithm \citep{emcee} with a Gaussian likelihood function and uniform priors for the correlation scale length of $r_0\in[1, 30]~{\rm cMpc}\,h^{-1}$ and the slope $\gamma\in[1, 3]$ to fit our measurements. The median and $16$th and $84$th percentiles of the posterior distribution define our the best estimate. 

With both the slope and scale length as free model parameters, we obtain $r_0^{\rm QG}=12.4^{+3.8}_{-2.3}~{\rm cMpc}\,h^{-1}$ and $\gamma_{\rm QG}=1.6^{+0.2}_{-0.3}$ as the best fits for the quasar-galaxy cross-correlation function. When fixing the slope to $\gamma_{\rm QG}=2.0$ to enable a comparison to other studies (see \S~\ref{sec:discussion}), we obtain $r_0^{\rm QG}=9.1^{+0.5}_{-0.6}~{\rm cMpc}\,h^{-1}$. 

Similarly, for the galaxy-galaxy auto-correlation, we obtain $r_0^{\rm GG}=3.9\pm0.4~{\rm cMpc}\,h^{-1}$ and $\gamma_{\rm GG}=1.9\pm0.1$ when fitting for both scale length and slope. When fixing the slope to $\gamma_{\rm GG}=1.8$ as in previous studies \citep[e.g.][]{GarciaVergara2017}, we obtain $r_0^{\rm GG}=4.1\pm0.3~{\rm cMpc}\,h^{-1}$. 


\subsection{Jointly Fitting both Correlation Functions}\label{sec:fit}

We can now attempt to study the clustering properties of the quasars themselves and infer the real-space quasar-quasar auto-correlation function $\xi_{\rm QQ}$ by means of $\xi_{\rm QG}$ and $\xi_{\rm GG}$. To this end, we follow a deterministic bias model \citep[see e.g.][]{Croom2001, GarciaVergara2017} 
and denote the density contrast of galaxies and quasars as $\delta_{\rm Q}$ and $\delta_{\rm G}$. Thus, the cross-correlation between quasars and galaxies is defined as $\xi_{\rm QG}=\langle \delta_{\rm Q}\delta_{\rm G}\rangle$. Assuming that galaxies and quasars trace the same underlying dark matter distribution, the galaxy and quasar density contrast can be expressed as $\delta_{\rm G}=b_{\rm G}(\delta_{\rm DM})\delta_{\rm DM}$ and $\delta_{\rm Q}=b_{\rm Q}(\delta_{\rm DM})\delta_{\rm DM}$, respectively, where $b_{\rm G}(\delta_{\rm DM})$ and $b_{\rm Q}(\delta_{\rm DM})$ denote the galaxy and quasar bias, which are (possibly non-linear) functions of the dark matter density contrast $\delta_{\rm DM}$. 
If we describe $\delta_{\rm Q}$ and $\delta_{\rm G}$ as two stochastic processes, their cross-correlation coefficient $\rho$ can be expressed as 
\begin{equation}
    \rho=\frac{\langle\delta_{\rm Q}\delta_{\rm G}\rangle}{\sqrt{\langle\delta_{\rm Q}\delta_{\rm Q}\rangle\langle\delta_{\rm G}\delta_{\rm G}\rangle}}. 
\end{equation}
However, given the deterministic relations for $\delta_{\rm Q}$ and $\delta_{\rm G}$ as defined above, it follows that $\rho=1$. Thus, the quasar-quasar auto-correlation $\xi_{\rm QQ}$ can be expressed as a combination of the other two correlation functions, $\xi_{\rm QG}$ and $\xi_{\rm GG}$, i.e.\ 
\begin{equation}
    \xi_{\rm QQ} = \xi_{\rm QG}^2 / \xi_{\rm GG}. \label{eq:xi_QQ}
\end{equation}

Analogously to Eqn.~\ref{eq:pl}, we parameterize all auto- and cross-correlation functions with a power-law, but now perform a joint fit to both the auto- and cross-correlation measurements. To this end, we fit $\xi_{\rm GG}$ as a function of $\gamma_{\rm GG}$ and $r_{0}^{\rm GG}$, as well as $\xi_{\rm QG}$ parameterized by $\gamma_{\rm QQ}, \gamma_{\rm GG}, r_{0}^{\rm QQ}$ and $r_{0}^{\rm GG}$ using Eqn.~\ref{eq:xi_QQ}. 

To enable a comparison to previous work (see \S~\ref{sec:comp}), we keep the slope of the two auto-correlation functions fixed, i.e.\ $\gamma_{\rm GG}=1.8$ \citep{GarciaVergara2017}, and $\gamma_{\rm QQ}=2.0$ \citep{Shen2007, Eftekharzadeh2015}, and only fit the two remaining free parameters, i.e.\ the scale lengths $r_{0}^{\rm GG}$ and $r_{0}^{\rm QQ}$. Our best estimates determined by the median and $16$th and $84$th percentile of the posterior probability distributions for the scale lengths of the galaxy auto-correlation function at $\langle z\rangle \approx 6.25$ is $r_{0}^{\rm GG}=4.1\pm0.3\,{\rm cMpc}\,h^{-1}$, while we obtain $r_{0}^{\rm QQ}=22.0^{+3.0}_{-2.9}\,{\rm cMpc}\,h^{-1}$ for the scale length of the quasars' auto-correlation function. These estimates for the power-laws and our measurements are shown in Fig.~\ref{fig:acf_ccf}. 

We also attempt to jointly fit for all four model parameters, i.e.\ two slopes and two correlation scale lengths. We caution that the parameters are strongly degenerate and approach the prior boundaries, but the marginalized posteriors suggest $r_{0}^{\rm GG}=4.0^{+0.4}_{-0.3}\,{\rm cMpc}\,h^{-1}$, $r_{0}^{\rm QQ}=23.7^{+4.3}_{-5.6}\,{\rm cMpc}\,h^{-1}$, $\gamma_{\rm GG}=1.9\pm0.1$ and $\gamma_{\rm QQ}=1.9\pm0.2$. 


\section{Mass estimates of the Quasars' Host Dark Matter Halos}\label{sec:DM}

In this section we will estimate the mass of the dark matter halos that harbor luminous quasars at $z\gtrsim 6$ based on the quasars' clustering properties. Our cosmological model $\Lambda$CDM provides a direct link between the clustering strength of a population of objects and the characteristic mass of their host dark matter halos, since more massive halos have a higher clustering bias \citep[e.g.][]{Kaiser1984, MoWhite1996, Tinker2010}. 

We will present three different approaches to constrain the host dark matter halo masses, all based on dark-matter-only simulations. We use N-body cosmological simulations to leverage the accurate representation of the halo clustering properties that they provide. Accurately modeling the distribution of halos in the universe is particularly important at the small scales probed by our observations (i.e. at distances of $r\lesssim10\,{\rm cMpc}\,h^{-1}$), as these scales are highly non-linear in overdense environments. Alternative approaches based on non-linear analytical extensions of linear theory to small scales \citep[such as the ``halo model'' framework, e.g.,][]{Cooray2001} 
would likely return results that are not accurate enough for our purposes. 


We first use a new N-body simulation \textsc{Flamingo-10k}, which is part of the \textsc{Flamingo} project \citep[][see \S~\ref{sec:flamingos}]{Schaye2023}, and assume simple ``step-function'' models for the halo occupation distributions \citep[HODs; e.g.][]{White2008} of quasars and galaxies. We then also compare to two empirical models for galaxies and SMBHs based on the \textit{Uchuu} N-body simulation \citep{Ishiyama2021}, namely the \textsc{UniverseMachine} \citep[][see \S~\ref{sec:UMac}]{Behroozi2019}, as well as the \textsc{Trinity} model \citep[][see \S~\ref{sec:trinity}]{HZhang2023a}.

\subsection{Host dark matter halo masses based on the \textsc{Flamingo-10k} cosmological simulation}\label{sec:flamingos}

As a first approach, we use the N-body cosmological simulation \textsc{Flamingo-10k} to directly infer the host halo masses of quasars and \oiii-emitters by matching the observed quasar-galaxy cross-correlation and galaxy-galaxy auto-correlation measurements with the clustering properties of dark matter halos.
\textsc{Flamingo-10k} is a new dark-matter-only simulation part of the \textsc{Flamingo} suite \citep{Schaye2023}, which makes use of the open-source code \textsc{Swift} \citep{Schaller2023}, a highly-parallel gravity and smoothed particle hydrodynamics solver. The simulation was run with $10080^3$ cold dark matter particles and $5600^3$ neutrino particles, and has the same volume as the \textsc{Flamingo} flagship hydrodynamical run ($2.8\,\rm cGpc$ on a side) but eight times more dark matter and neutrino particles, i.e.\ with a dark matter particle mass of $8.4\times10^8\,M_\odot$. 
The run adopts the ``3x2pt + all'' cosmology from \citet{abbott_des2022} ($\Omega_\mathrm{m} = 0.306$, $\Omega_\mathrm{b} = 0.0486$, $\sigma_8 = 0.807$, $\mathrm{H}_0 = 68.1\,\rm km\,s^{-1}\,{\rm Mpc}^{-1}$, $n_\mathrm{s} = 0.967$), with a summed neutrino mass of $0.06\,\mathrm{eV}$. Based on this simulation, we can model the clustering properties of any subset of $z\approx6$ dark matter halos in the mass range $M=10^{10.5}-10^{13}\,M_\odot$. Our simulation covers a total of $\approx10^8$ dark matter halos over this mass range, $60,857$ ($947$, $4$) of which above a halo mass of $10^{12}\,M_\odot$ ($10^{12.5}\,M_\odot$, $10^{13}\,M_\odot$). Thus, the number of lower-mass halos hosting \oiii-emitters is very large, but we need to ensure that our simulations include sufficient high-mass halos to limit the effects of cosmic variance. Assuming that $\sim 10-20$ of such halos are sufficient, the cross-correlation function estimated from the simulations will be independent of sampling effects due to cosmic variance for halo masses up to $\sim 10^{12.8-12.9}\,M_\odot$. For details on the modeling procedure and the simulation, we refer the reader to \citep{Pizzati2023, Pizzati2024}.

Assuming a prescription for how quasars and \oiii-emitters populate dark matter halos, we model the quasar-galaxy cross-correlation, $\chi_{\rm QG}$, as well as the galaxy-galaxy auto-correlation, $\chi_{\rm GG}$. Here, we assume a ``step-function'' HOD model for both quasars and \oiii-emitters, where halos are populated by quasars (\oiii-emitters) only above a minimum threshold mass $M_{\rm halo, min}$ ($M_{\rm halo, min,}^{\rm[OIII]}$). The two minimum host halo masses, $M_{\rm halo, min}$ and $M_{\rm halo, min,}^{\rm[OIII]}$, are free parameters in our model, which we infer by jointly fitting the observed correlation functions, $\chi_{\rm QG}$ and $\chi_{\rm GG}$.
We assume that these two correlation functions are independent, and write down a 2-d Gaussian likelihood for the parameters $M_{\rm halo, min}$ and $M_{\rm halo, min,}^{\rm[OIII]}$. By computing the median and 16th-84th percentiles of the marginalized likelihood distributions, we infer the minimum host dark matter halo mass of quasars to be $\log_{10}(M_{\rm halo, min}/M_\odot) = 12.43^{+0.13}_{-0.15}$, while we find that \oiii-emitters are hosted in halos of mass $\log_{10}(M_{\rm halo, min}^{\rm [OIII]}/M_\odot) = 10.56^{+0.05}_{-0.03}$. 
%
%
In a companion paper \citep{Pizzati2024}, we model the full mass distribution of the dark matter halos hosting quasars and galaxies by jointly fitting the clustering properties as well as the luminosity functions of quasars and \oiii-emitters, and find results that are in excellent agreement with the model presented here.  

\begin{figure*}[!t]
    \centering
    \includegraphics[width=0.49\textwidth]{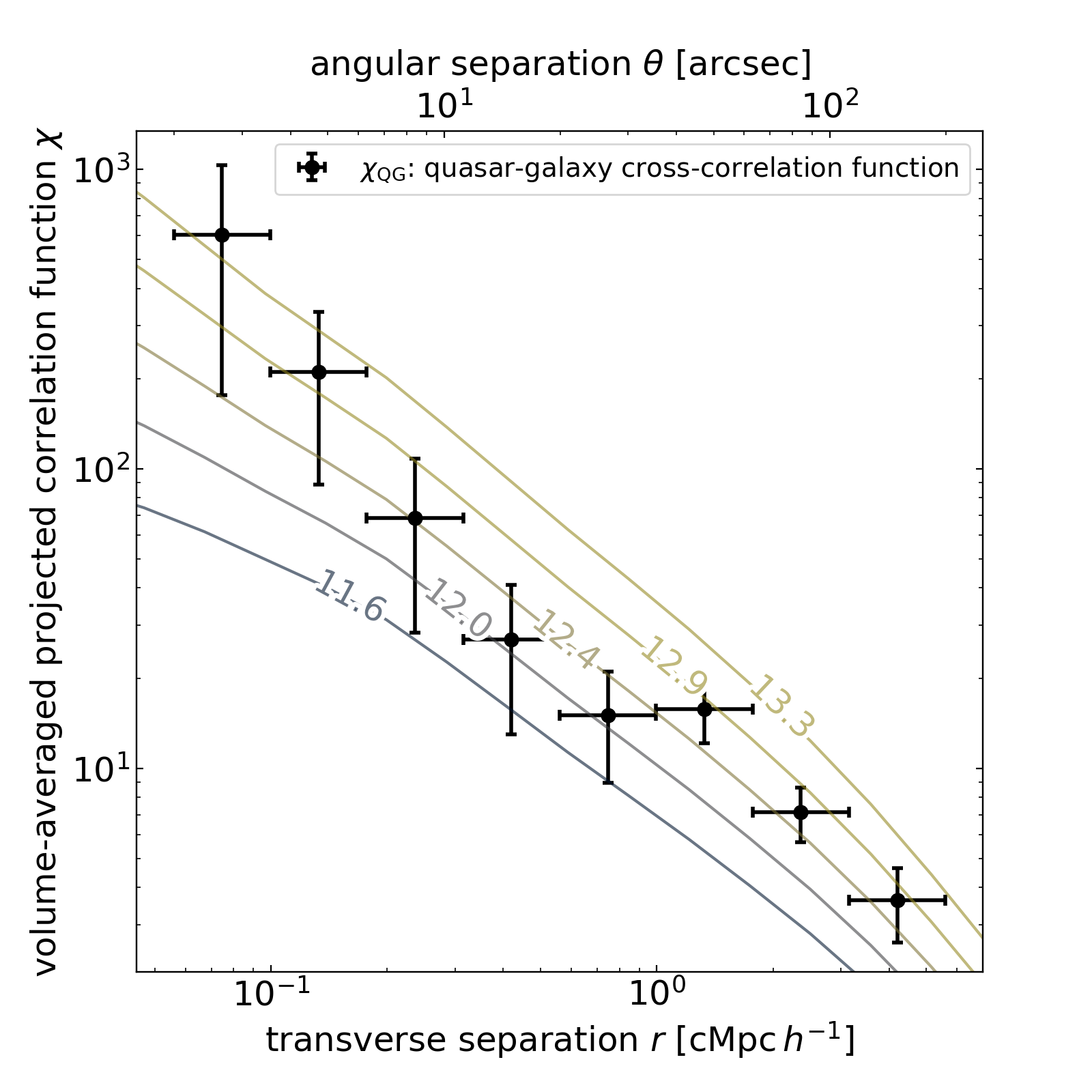}
    \includegraphics[width=0.49\textwidth]{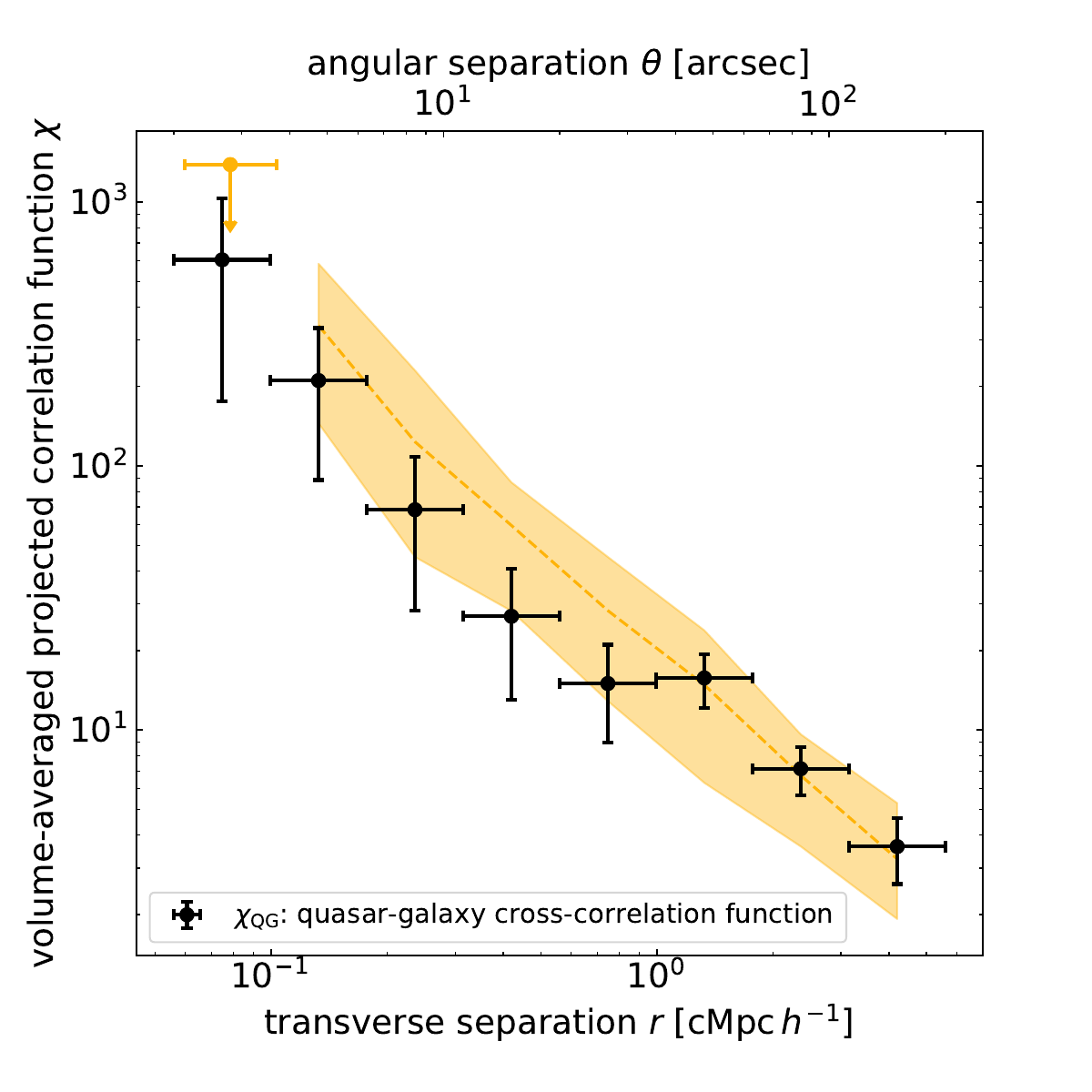}
    \caption{\textit{Left:} Comparison of the quasar-galaxy cross-correlation 
    measurement to models from the \textsc{Flamingo-10k} simulation for different minimum quasar host dark matter halo masses, $M_{\rm halo, min}$. By comparing the correlation functions of dark matter halos from the simulations to the measured quasar-galaxy cross-correlation and galaxy-galaxy auto-correlation, we obtain a best estimate for the quasars' minimum host halo mass of $\log_{10}(M_{\rm halo, min}/M_\odot)=12.43^{+0.13}_{-0.15}$. \textit{Right:} Comparison between the measurements of the quasar-galaxy cross-correlation with predictions (median: dashed yellow curve; $16$th-$84$th percentile: shaded region) for the observed quasar fields from the \textsc{Trinity} model, which imply an average host dark matter halo mass of $\log_{10} (M_\mathrm{halo}/M_\odot) = 12.14 ^{+0.24}_{-0.26}$.}
    \label{fig:flamingos}
\end{figure*}

In the left panel of Fig.~\ref{fig:flamingos}, we compare the predictions from the \textsc{Flamingo-10k} simulation with the observed quasar-galaxy cross-correlation measurements, for different values of the minimum halo mass for the quasar hosts, $M_\mathrm{halo, min}$. The cross-correlation terms are obtained by assuming a minimum host halo mass for \oiii-emitters equal to the median value determined above, i.e., $\log_{10}(M_{\rm halo, min}^{\rm [OIII]}/M_\odot) = 10.56$. 
The best fit between the simulated and observed cross- and auto-correlation functions is obtained with a minimum quasar host halo mass of $\log_{10}(M_{\rm halo, min}/M_\odot) \approx 12.43$. 

Note that the slope of the cross-correlation functions at very small scales, i.e. $r\lesssim 200~{\rm ckpc}\,h^{-1}$, appears to differ between the observations and the simulations. This could either be due to small number statistics given our current data set, as we only observe a total of $5$ galaxies for the inner most two data points, or alternatively, uncertainties in the modeling procedure could cause the discrepancy. For instance, the identification of subhalos becomes increasingly challenging at such small separations to a massive structure, and could thus alternate the shape of the correlation functions at small scales. However, our conclusions are not affected by the two inner most data points, as the uncertainties on these measurements are high and thus do not influence the best fit significantly. 

Additional differences at the smallest scales arise since we do not apply the same merging procedure to the subhalos in the simulations that we apply to the detected \oiii-emitters as described in \S~\ref{sec:grouping}. Merging the subhalos would lead to a slightly shallower slope in the simulated cross-correlation function at the smallest scales of $\lesssim 3 \arcsec$, but since the uncertainties on our measurements on the smallest scales are large due to the low number of detected galaxies, this has a negligible effect on our estimates. When applying the same merging procedure to subhalos in the \textit{Uchuu} simulations (see \S~\ref{sec:UMac}) as a check, we found that $\sim 2\%$ of the subhalos were affected compared to the $\sim13\%$ of \oiii-emitters affected by merging, which indicates that the \oiii-emitters show mostly clumpy substructure \textit{within} an individual subhalo. 
\\

In the following, we will briefly discuss two other methods to determine the host dark matter halo masses from the quasars' clustering properties to ensure consistency between different simulations and models, but moving forward, we will consider the measurement obtained here as our best estimate 
for the minimum host dark matter halo mass of $z\gtrsim 6$ quasars.

\subsection{Host dark matter halo masses based on the \textsc{UniverseMachine}}\label{sec:UMac}

A second method to determine the dark matter halo masses of the observed quasars is based on finding analogous systems to the observed quasar environments using the \textsc{UniverseMachine} \citep{Behroozi2019} applied to the \textit{Uchuu} N-body simulations. The details of this approach will be presented in detail in a forthcoming companion paper (Mackenzie et al.\ in prep.). In short, by comparing the number of simulated \oiii-emitters around halos of different masses to the number of companion galaxies observed in the quasar fields, a distribution of possible host dark matter halo masses can be determined for each observed quasar. 

We first construct a catalog of simulated environments around each halo, and select all halos above $10^{10}\,h^{-1}M_\odot$ (at $z=6.35$). In total this simulation contains $\sim10^{8.5}$ such halos, approximately $1.3\times 10^6$ ($24,525$, $1,860$, $85$, $8$) of which are above a mass threshold of $10^{11.5}\,M_\odot$ ($10^{12.1}\,M_\odot$, $10^{12.4}\,M_\odot$, $10^{12.7}\,M_\odot$, $10^{12.9}\,M_\odot$). 
Each halo has its own realisation where the halo position becomes the center of a mock observation (i.e. at the quasar position). The surrounding galaxies are then forward modeled through our completeness function and survey geometry. 
By calculating the joint distribution of the number of detected neighbouring galaxies vs.\ halo mass, and applying Bayes equation with a flat prior on halo mass, one can estimate the mass probability distribution for an observed number of \oiii-emitting neighbours. 
The resulting most probable dark matter halo masses range between $10^{10.85}\,M_\odot\lesssim M_{\rm halo}\lesssim 10^{12.85}\,M_\odot$ with a \textit{median} of $\langle M_{\rm halo}\rangle=10^{12.4\pm0.5}\,M_\odot$, where the wide range of host halo masses is largely driven by the cosmic variance observed between the different quasar fields. 
The estimate for the minimum host dark matter halo masses using constraints from the \textsc{UniverseMachine} are well consistent with the estimate based on the \textsc{Flamingo-10k} simulations. 



\subsection{Host dark matter halo masses based on \textsc{Trinity}}\label{sec:trinity}

Finally, we compare the observed $\chi_\mathrm{QG}$ with the prediction from the \textsc{Trinity} model \citep{HZhang2023a, HZhang2023b} of the halo--galaxy--SMBH connection, to infer the typical host dark matter halo mass of the quasars. Briefly, \textsc{Trinity} reconstructs empirical galaxy and black hole mass growth rate distributions as functions of halo mass from redshifts $z=0-10$, and forward models observables such as galaxy mass functions and quasar luminosity functions. SMBH properties such as radiative efficiencies, Eddington ratio distributions, and duty cycles are parameterized and constrained by observational data. 
For details on the \textsc{Trinity} implementation, we refer the reader to \citet{HZhang2023a}.

We utilize the \textsc{Trinity} mock catalogs of galaxies and SMBHs, with position, velocity, halo mass, galaxy mass, star formation rate, rest-frame UV luminosity, as well as black hole mass and bolometric luminosity (if hosting a SMBH) for each galaxy. These mock catalogs are generated also from the \textit{Uchuu} N-body simulation (box size: 2 cGpc $h^{-1}$, mass resolution: $3.27\times 10^8 M_\odot~h^{-1}$, \citealt{Ishiyama2021}), and will be made publicly available in a forthcoming \textsc{Trinity} publication (Zhang et al. in prep.).  
In order to calculate $\chi_\mathrm{QG}$ and host halo mass distributions, we use all quasars in our $z=6.35$ mock catalog, and assign each mock quasar a weight to quantify its degree of similarity to the here analyzed quasars. Specifically, the weight $w$ for a simulated quasar with black hole mass and bolometric luminosity $(M_\mathrm{BH}, L_\mathrm{bol})$ is a sum of the log-normal probabilities given the observed mass and luminosities, $\{(M_{\mathrm{BH},i}, L_{\mathrm{bol},i})\}_{i=1}^{N=4}$, of the EIGER quasars, i.e.
\begin{align}
    w = &\sum_{i=1}^{N=4} P\left[(M_\mathrm{BH}, L_\mathrm{bol}) | (M_{\mathrm{BH},i}, L_{\mathrm{bol},i})\right]\nonumber\\
        =&\sum_{i=1}^{N=4} \frac{1}{2\pi \sigma_{\mathrm{L_\mathrm{bol}},i} \sigma_{M_\mathrm{BH},i}}\nonumber \\
                                &\times \exp{\left[-\frac{(\log M_\mathrm{BH} - \log M_{\mathrm{BH},i})^2}{2\sigma_{M_{\mathrm{BH},i}}^2}\right]}\nonumber\\
                                &\times \exp{\left[-\frac{(\log L_\mathrm{bol} - \log L_{\mathrm{bol},i})^2}{2\sigma_{\mathrm{L_\mathrm{bol}},i}^2}\right]}\ ,
\label{eq:gauss_weights}
\end{align}
where $\sigma_{M_{\mathrm{BH},i}}$ and $\sigma_{\mathrm{L_\mathrm{bol}},i}$ are the measurement uncertainties in SMBH mass and luminosity of the $i$th quasar, respectively. For $\sigma_{M_{\mathrm{BH},i}}$, we also add a scatter of 0.5 dex in quadrature to account for the random scatter in virial estimates of $M_\mathrm{BH}$ \citep{McLure2002,Vestergaard2006}. 

When calculating the predicted $\chi_\mathrm{QG}$ we convert the galaxy UV luminosity, $L_\mathrm{UV}$, into $L_{\rm[O\,III], 5008}$ by applying the scaling relation from \citet{Matthee2023}, and include all mock galaxies with $L_{\rm[O\,III], 5008}\geq 10^{42}\,\rm erg\,s^{-1}$. To quantify $\chi_\mathrm{QG}$ uncertainties from \textsc{Trinity}, we simulate the observation 3000 times, i.e. we choose four mock quasars for every observation, where each quasar is randomly selected within the $1.5\sigma$ range of each observed quasar in the $(\log M_\mathrm{BH}, \log L_\mathrm{bol})$ space. We then calculate the $16^{\mathrm{th}}-84^{\mathrm{th}}$ percentile range of the measured $\chi_\mathrm{QG}$'s as the uncertainties from \textsc{Trinity}.  

As shown in right panel of Fig.~\ref{fig:flamingos}, the resulting $\chi_\mathrm{QG}$ from \textsc{Trinity} are consistent with the observed values within the measurement uncertainties. This consistency allows us to infer the host dark matter halo masses of the quasars by examining the host halo mass distribution of the mock quasars, weighted by the weights defined in Eqn.~\ref{eq:gauss_weights}. From this exercise, we obtain a median (with $16^{\mathrm{th}}-84^{\mathrm{th}}$ percentile) host halo mass of $\log_{10} (M_\mathrm{halo}/M_\odot) = 12.14 ^{+0.24}_{-0.26}$. 

\section{Estimating the Quasars' Duty Cycle and Lifetime}\label{sec:duty_cycle}

Our cosmological model $\Lambda$CDM allows us to determine the number density of the quasars' host dark matter halos $n_{\rm hosts}$ from their characteristic mass determined in the previous section \citep[e.g.][]{MoWhite1996, Tinker2010}, which in turn enables constraints on the quasars' duty cycle and lifetime. As quasars temporarily subsample their hosts, their number density $n_{\rm Q}$ can be expressed as
\begin{align}
    n_{\rm Q} \simeq \frac{t_{\rm Q}}{t_{\rm H}(z)} n_{\rm hosts}, \label{eq:duty}
\end{align}
where $t_{\rm Q}$ is the quasars' UV-luminous lifetime, and $t_{\rm H}(z)$ denotes the Hubble time at redshift $z$, and hence $f_{\rm duty}=t_{\rm Q}/t_{\rm H}(z)$ is the quasars' duty cycle at the given redshift \citep{MartiniWeinberg2001, HaimanHui2001}. We will now determine these timescales of quasar activity, which are essential for understanding the concomitant growth phases of early SMBHs. 

Using the \textsc{Flamingo-10k} simulation we find that the number density of halos with a mass larger than $\log_{10}(M_{\rm halo, min}/M_\odot) = 12.43^{+0.13}_{-0.15}$ is $n_{\rm host}=5.3^{+16.7}_{-3.9}\times 10^{-8}\rm\, cMpc^{-3}$. For estimating the number density of luminous quasars we use the quasar luminosity function (QLF) at $\langle z\rangle\approx 6.25$ by \citet{Schindler2023} and integrate it over the magnitude range $-26.5>M_{1450}>-29.5$ bracketing the observed magnitudes of the analyzed quasars (see Tab.~\ref{tab:observations}), which results in a quasar number density of $n_{\rm Q}\approx 0.23\,\rm cGpc^{-3}$. Using Eqn.~\ref{eq:duty} the two number densities determine a 
duty cycle of $f_{\rm duty}= 0.4^{+1.2}_{-0.3}\%$, and hence a UV-luminous quasar lifetime of approximately $t_{\rm Q}\sim10^{6.5}$~yr at $\langle z\rangle=6.25$. 

However, we caution that these timescale estimates depend sensitively on the number density of the host dark matter halos given the characteristic minimum halo mass derived in \S~\ref{sec:DM}. Thus even relatively small differences in the host halo mass estimates of approximately $\pm0.25$~dex can lead to variations in the derived timescales of $\pm1$~dex, i.e.\ a host dark matter halo mass of $\log_{10}(M_{\rm halo, min}/M_\odot)\approx 12.65$ ($\log_{10}(M_{\rm halo, min}/M_\odot)\approx 12.18$) with a number density of a factor of $10$ lower (higher) than our best estimate would correspond to a UV-luminous quasar lifetime of $t_{\rm Q}\sim10^{7.5}$~yr ($t_{\rm Q}\sim10^{5.5}$~yr). And yet, even with this uncertainty, these inferred UV-luminous lifetime estimates represent a major departure from the expected long timescales of $t_{\rm Q}\sim10^9$~yr and $f_{\rm duty}\sim 1$, required to grow the quasars' SMBHs. In order to match the abundance of host dark matter halos to the abundance of luminous quasars to ensure a duty cycle of unity, the required host dark matter halo mass would have to be $\sim 10^{13}\,M_\odot$. 


\section{Discussion}\label{sec:discussion}

\subsection{Comparison to previous studies}\label{sec:comp}

\begin{figure}
    \centering
    \includegraphics[width=0.48\textwidth]{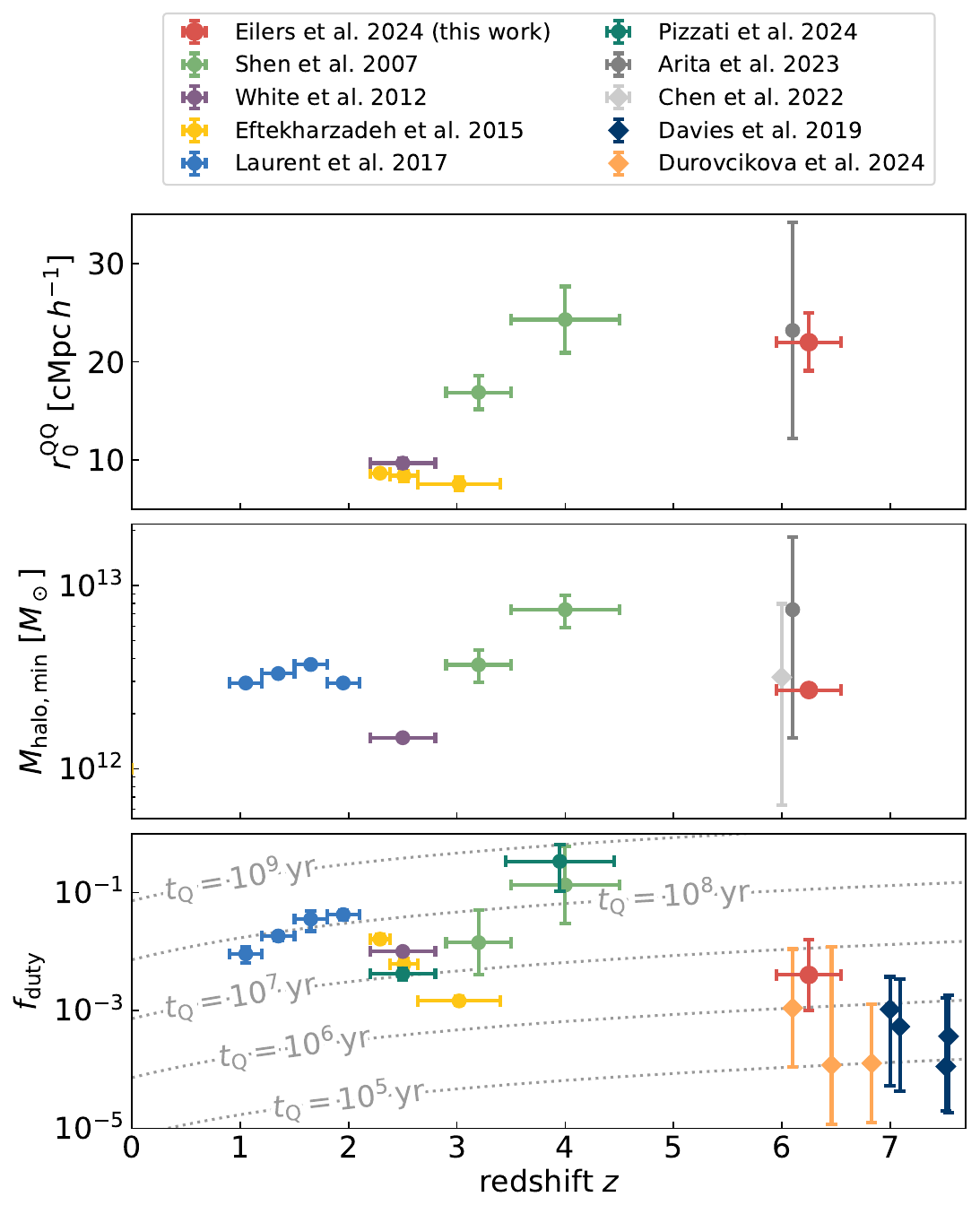}
    \caption{Redshift evolution of the quasars' auto-correlation scale length (\textit{top}), the minimum host dark matter halo mass (\textit{middle}), and the quasars' duty cycle (\textit{bottom}). Our inferred scale length is well consistent with the recent auto-correlation measurement by \citet{Arita2023}, but we infer a smaller host dark matter halo mass using dark matter only simulations (see \S~\ref{sec:DM}). Our results indicate a very small duty cycle of quasars at $z\gtrsim6$, which agrees with independent duty cycle measurements inferred from \lya\ damping wings \citep[diamond shaped data points;][]{Davies2019b, Durovcikova2024}. The host dark matter halo mass estimate by \citet{Chen2022} is also inferred from the transmitted flux along quasar sightlines, while all other measurements are derived from quasar clustering studies \citep[circular data points;][]{Shen2007, White2012, Eftekharzadeh2015, Laurent2017, Arita2023, Pizzati2023}. The grey dotted lines in the bottom panel indicate different quasar lifetimes. }
    \label{fig:compiliation}
\end{figure}

There is only one study to date measuring the auto-correlation function of $z\gtrsim6$ quasars, which is the recent result by the SHELLQs Collaboration \citep{Arita2023}. The authors analyze the auto-correlation function of approximately one hundred faint ($M_{1450}\gtrsim-25$) quasars at $z\sim6$ distributed over $891\,\rm deg^2$. They probe scales between $r\sim 10-1000\,{\rm cMpc}\,h^{-1}$, i.e.\ much larger scales than the scales of $r<10\,{\rm cMpc}\,h^{-1}$ we probe with our NIRCam WFSS data, and measure an auto-correlation scale length of $r_0^{\rm QQ}=23.7\pm11\, {\rm cMpc}\,h^{-1}$, which is in excellent agreement to our inferred estimate of $r_0^{\rm QQ}=22.0^{+3.0}_{-2.9}\, {\rm cMpc}\,h^{-1}$. They infer a host dark matter halo mass using the linear halo model software \texttt{HaloMod} \citep[][]{Murray2013, Murray2021}, and obtain a significantly larger host dark matter halo mass estimate of $\log_{10}(M_{\rm halo, min}/M_\odot)=12.9^{+0.4}_{-0.7}$. 

An independent approach to infer the host dark matter halos of a sample of luminous ($M_{\rm 1450}<-26.5$) quasars at $z\sim6$ with similar properties to the ones analyzed in this study is presented in \citet{Chen2022}. The authors use the flux transmission profile observed in the rest-frame UV quasar spectra to infer the density field around the quasars and obtain an estimate of the typical host dark matter halo masses of $\log_{10}(M_{\rm halo}/M_\odot)=12.5^{+0.4}_{-0.7}$, which is consistent with our results. 

In Fig.~\ref{fig:compiliation} we compare our results for the auto-correlation length, host dark matter halo mass and duty cycle to these previous analyses at $z\sim 6$. We also add measurements at lower redshifts to study the redshift evolution of the parameters. 
We confirm previous findings that the dark matter halo mass of quasars appears to be approximately constant throughout cosmic time, suggesting that there is a characteristic halo mass of a few times $M_{\rm halo}\sim 10^{12}\,M_\odot$ where quasars can be activated \citep[e.g.][]{Croom2005, Arita2023}. 

Interestingly, we know that galaxies typically quench at stellar masses $M_\star\sim 10^{11}\,M_\odot$ across a wide range of redshifts out to $z \sim 5$ \citep[e.g.][]{Peng2010, Behroozi2013, Caplar2015}, corresponding to a halo mass $M_{\rm halo}\sim 10^{12.0-12.5}\,M_\odot$, assuming an approximately constant stellar mass to halo mass ratio. Since there is considerable evidence that the quenching of galaxies is associated with, or possibly caused by, quasar activity \citep[e.g.][]{CancoDiaz2012, Caplar2018}, the identification of this same halo mass as the mass of the EIGER quasars' host halos at $z\sim6$ may not be unexpected. 



\subsection{Overdensities around high-redshift quasars?}\label{sec:diversity}

The numerous previous inconclusive results aiming to determine the existence of an overdense environment around high-redshift quasars have cast doubt on whether these quasars indeed trace the highest-density peaks in our universe. Our clustering measurement suggests that quasars reside in dark matter halos with a minimum mass of $\log_{10}(M_{\rm halo, min}/M_\odot)=12.43^{+0.13}_{-0.15}$, 
which implies that they do on average trace large overdensities but \textit{not necessarily} the rarest and highest density peaks. 

Our sample of four quasar fields has revealed a wide range of overdensities providing evidence for large cosmic variance. 
The observed density contrasts in the quasar fields range from $\delta(r<2\,{\rm cMpc})=65\pm15$ to $\delta(r<2\,{\rm cMpc})=3\pm4$. While J0148+0600 with $47$ \oiii-emitting galaxies in its surroundings likely resides in one of the highest-density peaks, and traces one of the largest high-redshift proto-cluster reported to date, 
J1030+0542 on the other hand appears to reside in an average density region with no enhancement compared to a random field \citep[which is in contrast to the proto-cluster reported in this field by][]{Mignoli2020}. 

It is interesting to note that the \cii\ emission arising from the molecular gas of the host galaxy of J1030+0542 was not detected with ALMA, indicating that the FIR luminosity of this quasar is significantly fainter than for most other $z\sim6$ quasars with similar rest-frame UV magnitudes \citep{Decarli2018}. This result, combined with the recent discoveries of large overdensities around sub-mm bright galaxies at $z>5$ \citep{Herard-Demanche2023, Sun2024}, provides circumstantial evidence that FIR luminosities might correlate more strongly with dark matter halo mass than the UV luminosities of high-redshift objects do.

\subsection{Implications for the early growth of SMBHs}

As the emission of quasar light is concomitant with black hole accretion, the timescales of quasar activity have implications for the growth of SMBHs. Previous constraints on the quasars' duty cycle based on clustering studies at redshifts of $1\lesssim z\lesssim 4$ suggest broadly an increase towards higher redshifts\footnote{We note that the increase in the quasars' duty cycle with redshift is not found by \citet{Eftekharzadeh2015}, who finds significantly lower duty cycles than \citet{Shen2007} at $z\sim3-4$. However, their quasar sample using BOSS includes much fainter objects, which makes the selection of quasars at these redshifts as well as the completeness evaluation of their sample more challenging.} as shown in the bottom panel of Fig.~\ref{fig:compiliation}, implying relatively long UV-luminous quasar lifetimes of $t_{\rm Q}\sim 10^7-10^8$~yr \citep[e.g.][]{Martini2004}. 

Our standard black hole growth model requires such long quasar lifetimes in order to explain the growth of the observed billion solar mass black holes of high-redshift quasars, and consequently high quasar duty cycles of $f_{\rm duty}\sim1$ have been postulated for quasars in the early universe \citep[e.g.][]{Martini2004, Volonteri2012}. However, the clustering analysis of $z\gtrsim 6$ quasars in this study surprisingly provides evidence for the contrary: we find a very small duty cycle, $f_{\rm duty}\ll 1$, indicating that galaxies do not shine as UV-luminous quasars throughout the whole Hubble time at these redshifts, thus posing additional challenges to our understanding of early SMBH growth. 

Interestingly, such low values are consistent with completely independent constraints on the UV-luminous duty cycle of quasars at similar redshifts derived from the extents of their proximity zones \citep[e.g.][]{Eilers2017a, Eilers2018b, Eilers2020, Andika2020, Morey2021, Satyavolu2023_rp} and damping wing features \citep[e.g.][]{Davies2019b, Durovcikova2024} that are observed in the quasars' rest-frame UV spectra. 
These studies estimate the number of ionizing photons emitted by a quasar that would results in the observed flux transmission profile along the line-of-sight, and thus enable constraints on their emission timescales for individual objects\footnote{Note that for quasars that are embedded in an already highly ionized IGM, with a neutral gas fraction of $x_{\rm HI}\sim 10^{-4}$ at $z\sim6$, the proximity zone size reflects the lifetime of the last quasar \textit{episode} and hence the duty cycle presents an upper limit on these episodic lifetime estimates. For quasars at $z\gtrsim 7$ that are embedded in a still fairly neutral IGM, $x_{\rm HI}\gtrsim 0.1$, the observed damping wing signature in the proximity zone allows for an estimate of the integrated quasar lifetime, i.e.\ the duty cycle, due to the long recombination time \citep[see e.g.][]{Davies2019b}. }. These observations have previously suggested UV-luminous quasar lifetimes of only $t_{\rm Q}\sim 10^6$~years, which imply similarly low duty cycles of $f_{\rm duty}\ll 1$ for high-redshift quasars \citep[see also][for a similar anlysis of \ion{He}{2} proximity zones at $z\sim4$]{Khrykin2019}. 

Thus, two independent arguments -- one based on the quasars' clustering properties and the other based on the observed flux transmission profiles in their spectra -- provide evidence for very short UV-luminous quasar activity timescales in the early universe. This suggests that accretion onto the SMBHs occurs in highly radiatively inefficient phases with a radiative efficiency $\epsilon$ of less than the fiducial $\sim10\%$ suggested by thin accretion disk models \citep[e.g.][]{Davies2019b}, or, alternatively, a significant fraction of the black hole growth happens in UV-obscured, dust-enshrouded environments \citep[e.g.][]{Satyavolu2023}. 

The latter is a particularly intriguing scenario in light of the recent discovery of an abundance of faint and highly dust-reddened active galactic nuclei (AGN) present at early cosmic times \citep{Harikane2023, Matthee2023b, Kocevski2023, Maiolino2023, Greene2023}, which might provide evidence for a highly obscured population of growing SMBHs. However, studies in the mid-infrared such as the recently published \textit{Systematic Mid-infrared Instrument Legacy Extragalactic Survey} \citep[SMILES;][]{Lyu2023} are necessary to identify completely dust-obscured growing SMBHs and securely determine the fraction of obscured quasars in the early universe. 
If black hole growth in completely dust-enshrouded, UV-obscured environments is indeed responsible for the discrepancy between the theoretically expected long black hole growth timescales and the observed short UV-luminous growth phases, we would expect $\sim100-1000$ dust-obscured quasars for every unobscured one, indicating the presence of a very large population of dust-obscured quasars in the early universe that we yet have to detect. While this would imply a very high obscuration fraction of $>99\%$, cosmological simulations \citep{Trebitsch2019, Ni2020, Vito2022, Bennett2024} as well as a recent study using ALMA observations \citep{Gilli2022} have shown that obscuration fraction of at least $\sim 80-90\%$ are expected for high-redshift quasars due to the presence of high-column density gas within the innermost regions of their host galaxies, resulting in low observable UV-luminous duty cycles.

\section{Summary}\label{sec:summary}

In this paper we present the first clustering measurement of luminous quasars at $z\gtrsim 6$ and \oiii-emitting galaxies in their environment in four quasar fields observed with JWST/NIRCam in imaging and WFSS mode as part of the EIGER project. We identify $522$ \oiii-emitting galaxy systems above a luminosity of $L_{\rm[O\,III], 5008}=10^{42}\,\rm erg\,s^{-1}$ in all five observed quasar fields, $92$ of which reside in close vicinity, i.e. $|\Delta v|\leq1000\,\rm km\,s^{-1}$, to the luminous quasars. 

We find a large diversity in the observed density contrast between the different quasar fields, ranging from density enhancements of $\delta(r<2\,{\rm cMpc})=65\pm15$ -- one of the most spectacular proto-clusters within the Epoch of Reionization discovered to date encompassing $47$ \oiii-emitting galaxies around the quasar J0148+0600 -- to $\delta(r<2\,{\rm cMpc})=3\pm4$, indicating no significant galaxy enhancement around the quasar J1030+0564. The other two fields around the super-luminous quasar J0100+2802 and the quasar J1148+5251, show density contrasts of $\delta(r<2\,{\rm cMpc})=29\pm10$ and $\delta(r<2\,{\rm cMpc})=16\pm8$, respectively. 

We present measurements of the volume-averaged projected galaxy-galaxy auto-correlation function and the quasar-galaxy cross-correlation function at an average redshift of $\langle z\rangle=6.25$. We parameterize the real-space correlation functions with a power-law and obtain estimates for the auto- and cross-correlation lengths of $r_0^{\rm GG}= 4.1\pm0.3 \,{\rm cMpc}\,h^{-1}$ and $r_0^{\rm QG}=9.1^{+0.5}_{-0.6}~{\rm cMpc}\,h^{-1}$ at $\langle z\rangle=6.25$, respectively (keeping the slopes $\gamma_{\rm GG}=1.8$ and $\gamma_{\rm QG}=2.0$ fixed). When jointly fitting the two correlation functions to infer the scale length of the quasar-quasar auto-correlation, we measure $r_0^{\rm QQ}= 22.0^{+3.0}_{-2.9} \,{\rm cMpc}\,h^{-1}$ (with fixed slope $\gamma_{\rm QQ}=2.0$). 

By comparing our measurements to dark matter only simulations and different empirical models, we find consistent estimates for the host dark matter halo mass: We determine a minimum host halo mass of $\log_{10}(M_{\rm halo, min}/M_\odot)=12.43^{+0.13}_{-0.15}$ as the best fit between our measurements and the cross-correlation function of dark matter halos in the \textsc{Flamingo-10k} simulation. When searching for analog systems to the quasar fields in \textsc{UniverseMachine} we find that such quasars are hosted in dark matter halos with average masses of $\log_{10}(M_{\rm halo}/M_\odot)=12.4\pm 0.5$. 
Lastly, when comparing the cross-correlation function to predictions from the \textsc{Trinity} model, we find host dark matter halo masses with a median of $\log_{10} (M_\mathrm{halo}/M_\odot) = 12.14^{+0.24}_{-0.26}$. All three estimates agree with each other within the uncertainties despite their very different approaches. 

These results for the mass of the quasars' host dark matter halos suggest that quasars do \textit{not necessarily} reside in the rarest, highest-density peaks, and could explain why some previous studies did not find large overdensities of galaxies around high-redshift quasars. It also indicates a smaller than expected duty cycle of quasars in the early universe, i.e.\ $f_{\rm duty}\ll1$. While the short UV-luminous duty cycles are consistent with independent results analyzing the flux transmission observed in rest-frame UV quasar spectra, the findings pose significant challenges for the growth of the early SMBHs which power the quasars' emission. Such short UV-luminous quasar lifetimes imply that (a) either the enormous $\gtrsim 10^9\,M_\odot$ black holes must have build up their mass extremely rapidly, e.g. via radiatively inefficient, ``super-Eddington'' accretion phases; or (b) a significant fraction ($\gtrsim99\%$) of the black hole growth happens in dust-enshrouded and UV-obscured environments, indicating the existence of a vast population of obscured quasars in the early universe. 

\begin{acknowledgements}
    The authors would like to thank the anonymous referee for thoughtful comments, which significantly improved our manuscript, and Jan-Torge Schindler, Jiamu Huang and Feige Wang for helpful discussions. 

    JFH and EP acknowledge support from the European Research Council (ERC) under the European Union’s Horizon 2020 research and innovation program (grant agreement No 885301). JM acknowledges support from the European Union (ERC, AGENTS, 101076224). 

    This work is based on observations made with the NASA/ESA/CSA James Webb Space Telescope. The JWST data presented in this article were obtained from the Mikulski Archive for Space Telescopes at the Space Telescope Science Institute, which is operated by the Association of Universities for Research in Astronomy, Inc., under NASA contract NAS 5-03127 for JWST. The specific observations analyzed are associated with program $\#1243$, and can be accessed via \dataset[DOI: 10.17909/m5mp-5v90]{https://doi.org/10.17909/m5mp-5v90}. 

    This work used the DiRAC Memory Intensive service (Cosma8) at the University of Durham, which is part of the STFC DiRAC HPC Facility (\url{www.dirac.ac.uk}). Access to DiRAC resources was granted through a Director’s Discretionary Time allocation in 2023/24, under the auspices of the UKRI-funded DiRAC Federation Project. The equipment was funded by BEIS capital funding via STFC capital grants ST/K00042X/1, ST/P002293/1, ST/R002371/1 and ST/S002502/1, Durham University and STFC operations grant ST/R000832/1. DiRAC is part of the National e-Infrastructure.

    We thank Instituto de Astrofisica de Andalucia (IAA-CSIC), Centro de Supercomputacion de Galicia (CESGA) and the Spanish academic and research network (RedIRIS) in Spain for hosting Uchuu DR1, DR2 and DR3 in the Skies \& Universes site for cosmological simulations. The Uchuu simulations were carried out on Aterui II supercomputer at Center for Computational Astrophysics, CfCA, of National Astronomical Observatory of Japan, and the K computer at the RIKEN Advanced Institute for Computational Science. The Uchuu Data Releases efforts have made use of the skun\@IAA\_RedIRIS and skun6\@IAA computer facilities managed by the IAA-CSIC in Spain (MICINN EU-Feder grant EQC2018-004366-P).

\end{acknowledgements}

\software{numpy \citep{numpy}, scipy \citep{scipy}, matplotlib \citep{matplotlib}, astropy \citep{astropy2013, astropy2018, astropy2022}, emcee \citep{emcee}}

\bibliography{literatur_hz}

\end{document}